\documentclass[lettersize,journal]{IEEEtran}
\usepackage[utf8]{inputenc}
\usepackage{amsmath,amsfonts}
\usepackage{algorithmic}
\usepackage{array}
\usepackage[caption=false,font=normalsize,labelfont=sf,textfont=sf]{subfig}
\usepackage{textcomp}
\usepackage{stfloats}
\usepackage{url}
\usepackage{verbatim}
\usepackage{graphicx}
\usepackage{booktabs}
\usepackage{amsmath}
\usepackage{float}
\usepackage[most]{tcolorbox}
\usepackage{enumitem}
\usepackage{tikz}
\usepackage[edges]{forest}
\usepackage{xcolor}
\usepackage{makecell}
\usepackage{balance}
\usepackage{CJKutf8}
\usepackage{multirow}
\usepackage{longtable}
\usepackage{booktabs}
\usepackage{multirow}
\usepackage{tabularx}
\usepackage{array}
\usepackage{booktabs}
\def\BibTeX{{\rm B\kern-.05em{\sc i\kern-.025em b}\kern-.08em
    T\kern-.1667em\lower.7ex\hbox{E}\kern-.125emX}}
\usepackage{balance}

\usepackage[table]{xcolor}
\usepackage{tabularx}
\usepackage{array}
\usepackage{booktabs}

\definecolor{taxblue}{RGB}{232,242,252}
\definecolor{taxbluehead}{RGB}{213,231,248}
\definecolor{taxorange}{RGB}{252,242,226}
\definecolor{taxorangehead}{RGB}{247,226,195}
\definecolor{taxgreen}{RGB}{235,247,232}
\definecolor{taxgreenhead}{RGB}{215,239,210}
\definecolor{taxgray}{RGB}{245,245,245}

\newcolumntype{C}[1]{>{\centering\arraybackslash}p{#1}}
\newcolumntype{Y}{>{\raggedright\arraybackslash}X}

\usepackage{cite}  
\begin{document}
\title{A Comprehensive Survey on Event Camera Representation Learning}
\author{Hongwei Ren*\dag, Youxin Jiang*, Tuopusen Huang, Lei Yu, Xiao Wang, Min Liu, Lin Wang, Xiangqian Wu
\thanks{* Equal contribution,\dag Corresponding author.
}}

\markboth{Journal of \LaTeX\ Class Files,~Vol.~18, No.~9, September~2020}%
{How to Use the IEEEtran \LaTeX \ Templates}

\maketitle

\begin{abstract}
Event cameras offer distinctive advantages, including microsecond-level latency and high dynamic range, rendering them promising for challenging perception tasks. 
Inspired by biological vision, they output asynchronous and sparse event streams rather than dense image frames, creating a fundamental mismatch with mainstream neural networks. 
This survey reviews recent advances in event camera representation learning from the perspective of converting raw event streams into learnable representations. 
We first organize existing methods according to whether they rely on a single principal event representation or jointly exploit multiple complementary representations. 
Single-representation methods are further categorized into dense-based representations, which regularize events into structured grid-like forms, and sparse-based representations, which preserve event-native discrete spatio-temporal structures. 
Multi-representation methods are organized into dense–dense and dense–sparse hybrid formulations that exploit complementary representation properties. 
This representation-centric taxonomy clarifies how different paradigms balance structural regularity, temporal fidelity, sparsity preservation, and architectural compatibility. 
For each paradigm, we examine the underlying design choices, modeling principles, and task-level implications. We further summarize standard benchmarks and evaluation settings across representative high-level perception and low-level vision tasks. Finally, we discuss open problems and outline future directions from fixed representation design toward adaptive representation optimization, improved fidelity–efficiency trade-offs, and more scalable event-based perception systems.
\end{abstract}
\section{Introduction}
With the growing demand for computer vision in high-speed and high-dynamic-range (HDR) scenarios, the limitations of traditional frame-based cameras have become increasingly evident.
Constrained by fixed sampling rates and limited dynamic range, these sensors often suffer from motion blur, limited exposure adaptability, and substantial data redundancy, making them inadequate for perception in extreme environments \cite{chakravarthi2024recent}.
In contrast, event cameras, inspired by biological vision \cite{gallego2020event}, offer a fundamentally different sensing paradigm through asynchronous sensing. 
Instead of capturing full frames at fixed intervals, they respond independently to pixel-wise brightness changes, providing microsecond-level latency, high dynamic range, sparse event streams, and low power consumption \cite{4444573}. 
As a result, event cameras have shown strong potential in demanding applications such as autonomous driving, robotics, and high-speed industrial inspection.
\begin{figure}[t]
    \centering
    \includegraphics[width=\columnwidth]{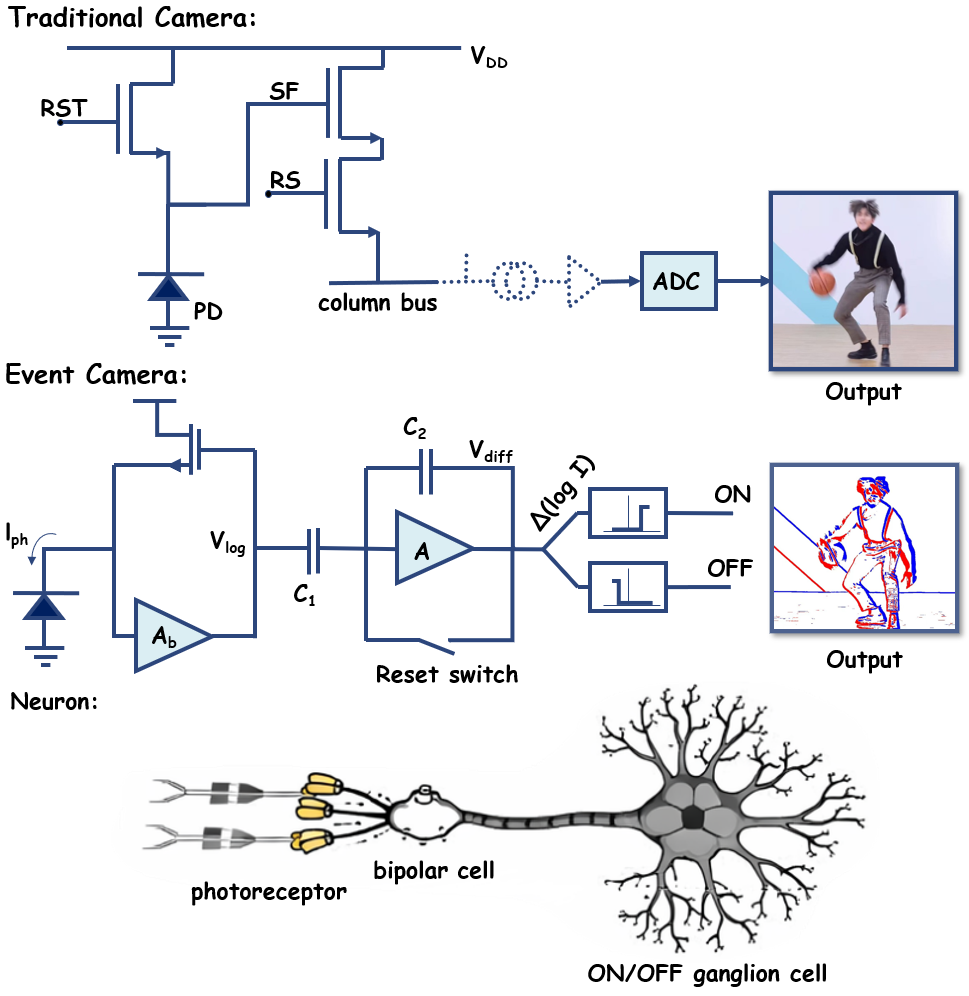}
    \caption{Comparison between traditional frame-based cameras, event cameras, and biological visual pathways. Event cameras asynchronously generate polarity-aware ON/OFF events in response to local brightness changes, resembling retinal ON/OFF signaling.}
    \label{fig:event_vs_traditional_bio}
\end{figure}

From a hardware perspective, event cameras are typically
implemented using custom-designed complementary metal-
oxide-semiconductor (CMOS) circuits~\cite{4444573}, in which each pixel operates independently as an event generator.
This pixel-level sensing mechanism is illustrated in
Fig.~\ref{fig:event_vs_traditional_bio}.
Each pixel contains a photodiode and a logarithmic amplifier that continuously monitors changes in log-intensity.
When the brightness change since the last event at that pixel exceeds a predefined contrast threshold, the local circuit is triggered to generate a new event with microsecond-level temporal precision.
This process can be formulated as follows:
\begin{equation}
    \Delta L(x_k,y_k,t_k)
    =
    L(x_k,y_k,t_k)-L(x_k,y_k,t_k^-),
\end{equation}
\begin{equation}
    |\Delta L(x_k, y_k, t_k)| \geq C,
    \quad
    p_k = \operatorname{sign}\big(\Delta L(x_k, y_k, t_k)\big),
\end{equation}
where \(k\) denotes the event index, \(t_k^-\) denotes the timestamp of the previous event at the same pixel, \(\Delta L\) is the accumulated log-intensity change, \(C\) is the contrast threshold, and \(p_k\in\{+1,-1\}\) denotes the event polarity.
The resulting output forms a continuous asynchronous event stream, represented as:
\begin{equation}
\mathcal{E} = \{ e_k \}_{k=1}^{N}, \quad e_k = (x_k, y_k, t_k, p_k),
\end{equation}
where \(e_k\) denotes the \(k\)-th event, \(N\) is the number of events in the stream, and \((x_k, y_k)\), \(t_k\), and \(p_k\) represent its spatial location, timestamp, and polarity, respectively.
\begin{figure*}[t]
    \centering
    \includegraphics[width=0.8\textwidth]{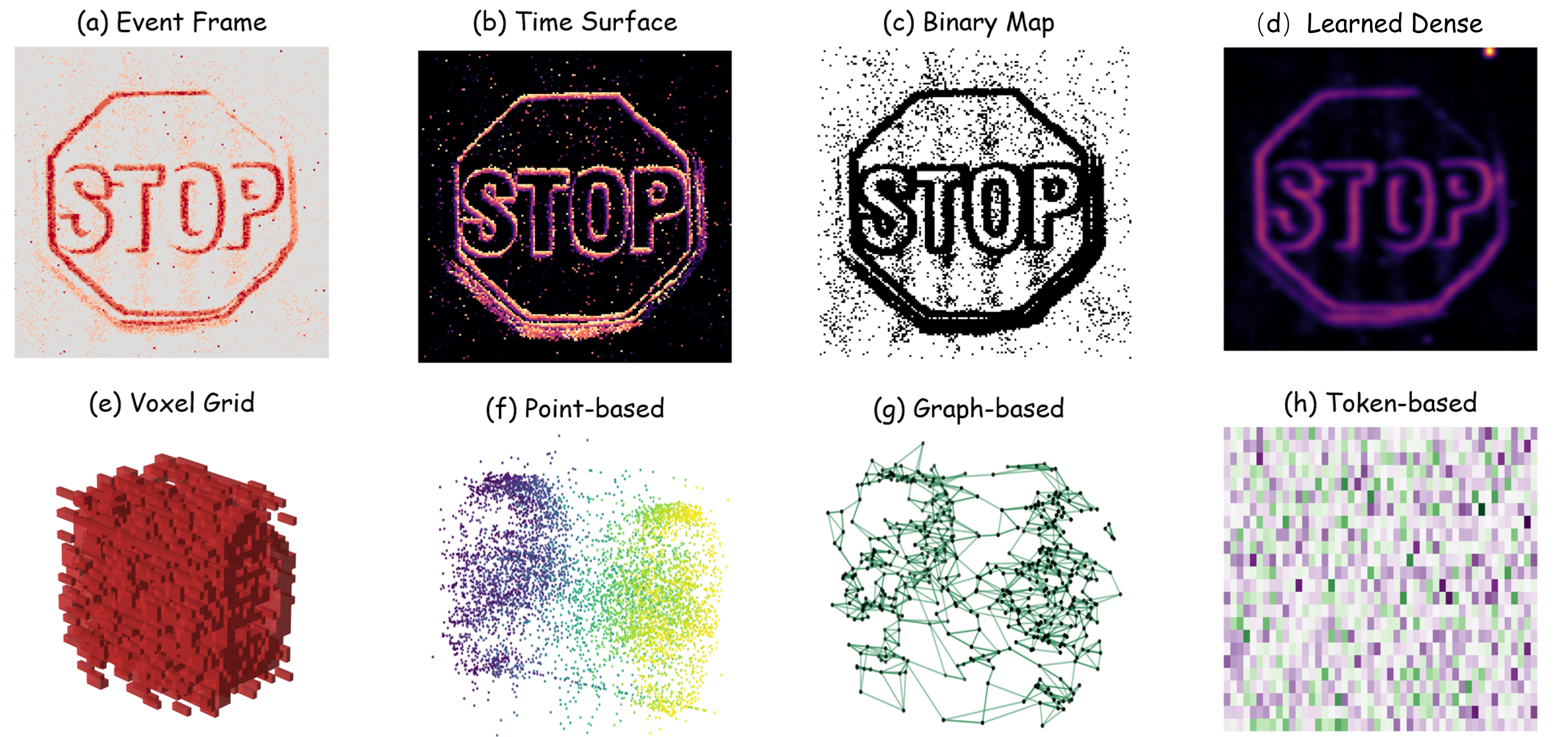} 
    \caption{Overview of diverse event camera representation paradigms using a ``STOP'' sign scene. (a), (b), (c), (e) show different frame-based mappings; (d) depicts end-to-end learned feature activations;(f) preserves raw sparsity in a 3D temporal point cloud; (g) illustrates graph-based connectivity; (h) represents a structured token latent space.}
    \label{fig:representations}
\end{figure*}

Although the event stream in Address-Event Representation (AER) form preserves fine-grained temporal information, it is fundamentally mismatched with mainstream neural networks.
Unlike standard image data, event streams are spatially sparse, temporally asynchronous, and structurally irregular.
Consequently, how to transform raw event streams into representations compatible with modern neural architectures has become a central challenge in event-based vision~\cite{gehrig2019end}.
To address this mismatch, a wide range of event representations have been developed.

In this survey, we first distinguish existing methods according to whether they rely on a \textbf{single principal event representation} or jointly exploit \textbf{multiple complementary representations}, as summarized in Fig.~\ref{fig:taxonomy_overview_complex}.
Single-representation methods are further categorized according to the structural form presented to the principal feature extractor into \underline{dense-based} and \underline{sparse-based} representations.
Dense-based representations regularize event streams into structured grid-like formats for compatibility with conventional neural architectures, whereas sparse-based representations preserve event-native discrete spatio-temporal structures for more faithful modeling of temporal dynamics and data sparsity.
Multi-representation methods, in contrast, explicitly construct and jointly exploit multiple event representations within the same perception pipeline to leverage their complementary structural and temporal characteristics.
For single-representation methods, the representation type is determined by the structural form presented to the principal feature extractor, rather than by whether the subsequent neural operations are implemented densely or sparsely.

\textbf{1) Single-representation methods:}
Single-representation methods rely on one principal event representation throughout the main perception pipeline.
According to the structural form presented to the principal feature extractor, they can be further divided into dense-based and sparse-based representations.
Representative single-representation structures are illustrated in Fig.~\ref{fig:representations}.

\tikzstyle{my-box}=[
rectangle,
draw=black,
rounded corners,
text opacity=1,
minimum height=1.5em,
minimum width=5em,
inner sep=2pt,
align=center,
fill opacity=.5,
]

\makeatletter
\@ifundefined{color}{}{
    \definecolor{typeblue}{RGB}{235,245,255}    
    \definecolor{evalgreen}{RGB}{240,255,240}   
    \definecolor{mitiorange}{RGB}{255,245,230}  
}
\makeatother

\newsavebox{\tempTreeBox}

\begin{figure*}[t]
\centering

\begingroup

\scriptsize
\setlength{\tabcolsep}{5pt}
\renewcommand{\arraystretch}{1.10}

\setlength{\arrayrulewidth}{0.45pt}
\arrayrulecolor{black!65}

\setlength{\fboxrule}{0.55pt}
\setlength{\fboxsep}{0pt}


\fcolorbox{black!80}{black!78}{%
    \parbox[c][0.64cm][c]{\textwidth}{%
        \centering
        \color{white}
        \normalsize\bfseries
        Event Camera Representation Learning
    }%
}

\vspace{1.4mm}


\fcolorbox{black!65}{gray!16}{%
    \parbox[c][0.56cm][c]{\textwidth}{%
        \centering
        \small\bfseries
        Single-Representation Methods
    }%
}

\vspace{1.0mm}


\begin{tabular}{|
    >{\centering\arraybackslash}m{0.20\textwidth}|
    >{\raggedright\arraybackslash}m{0.745\textwidth}|
}

\hline


\rowcolor{taxbluehead}
\multicolumn{2}{|l|}{%
    \rule{0pt}{3.0ex}
    \hspace{4pt}
    \small\bfseries
    Dense-based Methods
}
\\
\hline

\rowcolor{taxblue}
\textbf{\S~IV-A Event Frame}
&
Frame-to-Event~\cite{perez2013mapping},
Deep SNN~\cite{amir2017low},
SLAYER~\cite{shrestha2018slayer},
TA-SNN~\cite{yao2021temporal},
STSC-SNN~\cite{xu2023stsc},
SNN-BP~\cite{lee2016training},
MambaEVT~\cite{wang2025mambaevt},
HDETrack~\cite{wang2025event},
CRSOT~\cite{zhu2025crsot},
CMT-MDNet~\cite{wang2023visevent},
AsyNet~\cite{messikommer2020event},
SAM~\cite{liang2021event},
EV-SegNet~\cite{alonso2019evsegnet},
EventDAM~\cite{zhu2025depth},
STE-FlowNet~\cite{ding2022spatio},
CTN~\cite{zhao2022transformer},
RAMNet-fusion~\cite{gehrig2021ramnet},
RENet-fusion~\cite{zhou2023rgb},
CAFR~\cite{cao2024embracing},
HALSIE~\cite{das2024halsie}.
\\
\rowcolor{taxblue}
\multicolumn{2}{|c|}{\rule{0pt}{1.5pt}}
\\


\rowcolor{taxblue}
\textbf{\S~IV-B Time Surface}
&
EV-FlowNet~\cite{zhu2018ev},
SAE~\cite{benosman2013event},
HMAX-SNN~\cite{xiao2019hmax},
HMAX-SPA~\cite{liu2020event},
HATS~\cite{sironi2018hats},
ECSNet~\cite{chen2022ecsnet},
Feature tracking~\cite{mueggler2017fast}.
\\
\rowcolor{taxblue}
\multicolumn{2}{|c|}{\rule{0pt}{1.5pt}}
\\


\rowcolor{taxblue}
\textbf{\S~IV-C Binary Map}
&
Bina-Rep~\cite{barchid2022bina},
TBR~\cite{9412991},
Mambapupil~\cite{wang2024mambapupil},
TDTracker~\cite{huang2025exploring}.
\\
\rowcolor{taxblue}
\multicolumn{2}{|c|}{\rule{0pt}{1.5pt}}
\\


\rowcolor{taxblue}
\textbf{\S~IV-D Voxel Grid}
&
VMST-Net~\cite{liu2023voxel},
EVSTr~\cite{xie2024event},
ACE-BET~\cite{liu2022fast},
CES~\cite{lin2025compressed},
E2VID~\cite{rebecq2019high},
Sparse-E2VID~\cite{cadena2023sparse},
E2VID+~\cite{paredes2021back},
CEUTrack~\cite{tang2025revisiting},
AEC~\cite{peng2023better},
SRFNet~\cite{pan2024srfnet},
Taming-CM~\cite{shiba2023taming},
TimeReplayer~\cite{he2022timereplayer},
TimeLens~\cite{tulyakov2021timelens},
A2OF~\cite{wu2022video},
CBMNet-L~\cite{kim2023cbmnet},
E-VFI~\cite{liu2024video},
RVT-B~\cite{gehrig2023recurrent},
SAST~\cite{peng2024scene},
SAST-CB~\cite{peng2024scene},
SMamba~\cite{yang2025smamba},
SpikeFPN~\cite{zhang2024automotive},
MSRNN~\cite{zhang2023multi},
ECOSNet~\cite{zhu2024continuous},
ESS~\cite{sun2022ess},
EDCNet~\cite{zhang2021exploring},
Hybrid-Seg~\cite{li2025efficient},
MambaSeg~\cite{gu2026mambaseg},
RAMNet~\cite{gehrig2021ramnet},
E2Depth~\cite{hidalgo2020learning},
Distil-E2D~\cite{lee2025distile2d},
MDDE+~\cite{hidalgo2020learning},
ULODE~\cite{zhu2019unsupervised},
EMoDepth~\cite{zhu2023selfsupervised},
DCR-EFlow~\cite{sun2025dcr},
DCEIFlow~\cite{wan2022learning},
E-RAFT~\cite{gehrig2021eraft},
TMA~\cite{liu2023tma},
EVA-Flow~\cite{ye2025eva},
ADM-Flow~\cite{luo2023learning},
Spike-FlowNet~\cite{lee2020spikeflownet},
TORE~\cite{baldwin2022time},
EFNet~\cite{han2022event},
EIFNet~\cite{yang2023event},
MAENet~\cite{sun2024motion},
CFFNet~\cite{li2024coarse},
MAT~\cite{xu2025motion},
FEVD~\cite{kim2024frequency},
RED~\cite{perot2020learning},
FireNet~\cite{scheerlinck2020fast},
FireNet+~\cite{paredes2021back}.
\\
\rowcolor{taxblue}
\multicolumn{2}{|c|}{\rule{0pt}{1.5pt}}
\\


\rowcolor{taxblue}
\textbf{\S~IV-E Learned Dense}
&
EST~\cite{gehrig2019end},
ASTMNet \cite{li2022asynchronous},
Matrix-LSTM~\cite{cannici2020differentiable}.
\\
\hline
\end{tabular}



\begin{tabular}{|
    >{\centering\arraybackslash}m{0.20\textwidth}|
    >{\raggedright\arraybackslash}m{0.745\textwidth}|
}

\hline


\rowcolor{taxorangehead}
\multicolumn{2}{|l|}{%
    \rule{0pt}{3.0ex}
    \hspace{4pt}
    \small\bfseries
    Sparse-based Methods
}
\\
\hline


\rowcolor{taxorange}
\textbf{\S~V-A Token-based}
&
Event Transformer~\cite{sabater2022event},
Event Transformer$^{+}$~\cite{sabater2023event},
GET~\cite{peng2023get},
ETB~\cite{jiang2024token}.
\\
\rowcolor{taxorange}
\multicolumn{2}{|c|}{\rule{0pt}{1.5pt}}
\\


\rowcolor{taxorange}
\textbf{\S~V-B Graph-based}
&
RG-CNN~\cite{bi2020graph},
EV-VGCNN~\cite{deng2022voxel},
VMV-GCN~\cite{xie2022vmv},
SlideGCN~\cite{li2021graph},
AEGNN~\cite{schaefer2022aegnn},
EventMG~\cite{wu2025eventmg}.
\\
\rowcolor{taxorange}
\multicolumn{2}{|c|}{\rule{0pt}{1.5pt}}
\\


\rowcolor{taxorange}
\textbf{\S~V-C Point-based}
&
ST-EC~\cite{wang2019space},
EventNet~\cite{sekikawa2019eventnet},
REPC~\cite{chen2022efficient},
TTPOINT~\cite{ren2023ttpoint},
SpikePoint~\cite{ren2023spikepoint},
PEPNet~\cite{ren2024simple},
EventMamba~\cite{ren2024rethinking},
E2B~\cite{ren2025e2b},
HSPC~\cite{tang2025event},
SECNet~\cite{ren2026scalable}.
\\
\hline
\end{tabular}


\vspace{2.0mm}


\fcolorbox{black!65}{gray!16}{%
    \parbox[c][0.56cm][c]{\textwidth}{%
        \centering
        \small\bfseries
        Multi-Representation Methods
    }%
}

\vspace{1.0mm}


\begin{tabular}{|
    >{\centering\arraybackslash}m{0.20\textwidth}|
    >{\raggedright\arraybackslash}m{0.745\textwidth}|
}

\hline


\rowcolor{taxgreenhead}
\multicolumn{2}{|l|}{%
    \rule{0pt}{3.0ex}
    \hspace{4pt}
    \small\bfseries
    Hybrid-based Methods
}
\\
\hline


\rowcolor{taxgreen}
\textbf{\S~VI-A Dense--Dense Hybrid}
&
EV-ACT~\cite{gao2023action},
EFV~\cite{yuan2023learning},
Spike-EVPR~\cite{liu2026spike},
MTC~\cite{chen2019multi},
EFV++~\cite{chen2025retain},
MAD-Det~\cite{chen2025motion},
SMV-EAR~\cite{fan2026smv}.

\\
\rowcolor{taxgreen}
\multicolumn{2}{|c|}{\rule{0pt}{1.5pt}}
\\


\rowcolor{taxgreen}
\textbf{\S~VI-B Dense--Sparse Hybrid}
&
MTGA~\cite{zhang2025mtga},
EventCrab~\cite{cao2024eventcrab},
MTGNet~\cite{lin2026event}.
\\

\hline
\end{tabular}

\endgroup


\caption{
Taxonomy of event camera representation learning.
Existing methods are first categorized according to whether they rely
on a single principal event representation or jointly exploit multiple
complementary representations.
Single-representation methods are further divided into dense- and
sparse-based paradigms according to structural regularity and event
sparsity, while multi-representation methods are organized into
dense--dense and dense--sparse hybrid formulations.
}

\label{fig:taxonomy_overview_complex}

\end{figure*}

\textbf{a) Dense-based representations:} Dense-based representations convert raw event streams into regular grid-like structures, such as maps, tensors, or spatio-temporal volumes. The main motivation of this paradigm is to improve compatibility with mature deep learning architectures originally developed for dense visual inputs, such as CNNs and Transformers. 

By regularizing asynchronous events into structured arrays, these representations provide explicit spatial organization and enable efficient processing with conventional dense operators.
However, such regularization inevitably introduces temporal quantization and information aggregation, which may weaken the native sparsity and fine-grained temporal precision of event data.

\emph{\textbf{Map-based representations:}} This is one of the most widely used dense formulations. 
It converts the asynchronous event stream within a temporal window into a structured map representation through an aggregation function:
\begin{equation}
    \mathcal{M} = \Phi(\mathcal{E}_{[t_s,t_e]}), 
    \quad \mathcal{M} \in \mathbb{R}^{H \times W \times C},
\end{equation}
where \(\mathcal{M}\) denotes the resulting map-based representation,
\(\mathcal{E}_{[t_s,t_e]}\) denotes the set of events accumulated within the time interval \([t_s,t_e]\), \(t_s\) and \(t_e\) are the start and end timestamps of the interval, \(\Phi(\cdot)\) denotes the event aggregation operator, and \(H\), \(W\), and \(C\) denote the height, width, and channel number of the resulting map, respectively. Although structurally similar to conventional image tensors, this representation encodes temporal information through the aggregation process. Typical examples, as shown in Fig.~\ref{fig:representations}(a)--(c), include event frames, time surfaces, and binary maps.

\emph{\textbf{Voxel-grid representations:}} As an extension of map-based encoding to the spatio-temporal domain, voxel-grid representation is one of the most widely used dense event formulations. Given an event stream within a temporal window, voxel-grid representations organize events into a structured spatio-temporal tensor:
\begin{equation}
    \mathcal{V} = \{V_b\}_{b=1}^{B}, \quad
    \mathcal{V} \in \mathbb{R}^{B \times H \times W \times C},
\end{equation}
where \(B\) denotes the number of temporal bins, \(H\) and \(W\)
denote the height and width of the spatial grid, and \(C\) denotes
the number of channels. Each slice \(V_b\) aggregates the events
assigned to the \(b\)-th temporal bin based on predefined accumulation rules, such as event counting, polarity summation, or interpolated voting.
In this way, voxel-grid representations preserve temporal ordering more effectively than single-map encodings, while remaining compatible with standard dense neural architectures.

\emph{\textbf{Learned dense representations:}}
Instead of relying on manually designed accumulation rules, learned dense representations parameterize the event-to-grid mapping itself and construct explicit dense representations directly from raw event streams:
\begin{equation}
    \mathcal{Z} = f_{\theta}(\mathcal{E}), \quad
    \mathcal{Z} \in \mathbb{R}^{T \times H \times W \times C},
\end{equation}
where \(\mathcal{E}\) denotes the input event stream,
\(f_{\theta}(\cdot)\) is a trainable event-to-representation mapping parameterized by \(\theta\), and
\(\mathcal{Z}\) denotes the resulting dense representation.
Here, \(T\), \(H\), \(W\), and \(C\) denote the temporal dimension, height, width, and channel number, respectively.
Unlike handcrafted dense representations whose accumulation or discretization rules are predefined, the construction of \(\mathcal{Z}\) is optimized jointly with the downstream objective, allowing temporal projection and aggregation to adapt to the data and task supervision.

In this survey, we restrict learned dense representations to methods in which the event-to-grid mapping itself is trainable and produces an explicit dense representation before the principal feature extractor.
Learned hidden states or task-network features alone are therefore not regarded as learned event representations.

\textbf{b) Sparse-based representations:}
In contrast to dense-based representations, sparse-based representations preserve event sparsity without projecting the entire event stream onto a regular image-plane grid.
They organize asynchronous events as discrete tokens, relational graphs, or spatio-temporal point sets, while retaining data-dependent sparse structures.
This formulation better preserves fine-grained temporal information and avoids computation over large inactive regions.
However, sparse representations generally require specialized token interaction, neighborhood construction, graph propagation, or point aggregation mechanisms, which may introduce irregular memory access and additional structural processing overhead.

\emph{\textbf{Token-based representations:}}
To better exploit Transformer architectures, some methods decompose event observations or local event regions into token sequences:
\begin{equation}
\mathcal{T} = [\tau_1, \tau_2, \dots, \tau_L], \quad \mathcal{T} \in \mathbb{R}^{L \times D},
\end{equation}
where \(\tau_i\) denotes the \(i\)-th event token, \(L\) denotes the sequence length, and \(D\) denotes the embedding dimension.
This formulation enables attention-based sequence modeling to capture local and long-range spatio-temporal dependencies.
Although token sequences are structurally regular, their tokens are usually generated from sparse event subsets rather than dense image-plane grids, and are therefore categorized as sparse-based representations in this survey.

\emph{\textbf{ Graph-based representations:}}
Graph-based representations are a representative sparse event formulation. They represent the event stream as a graph:
\begin{equation}
    \mathcal{G} = (\mathcal{N}, \mathcal{A}, \mathbf{X}), \qquad
    \mathcal{A} = \{(n_i, n_j)\mid \phi(n_i, n_j)=1\},
\end{equation}
where \(\mathcal{N} = \{n_i\}_{i=1}^{M}\) denotes the node set, \(M\) denotes the number of nodes, \(\mathcal{A}\) denotes the edge set, and \(\mathbf{X} \in \mathbb{R}^{M \times D}\) is the node feature matrix.
The connectivity function \(\phi(\cdot,\cdot)\) determines whether two nodes are connected according to spatial, temporal, or semantic affinity. Each node may correspond to an individual event or a local spatiotemporal region. By organizing event data into a non-Euclidean relational structure, graph-based representations are well-suited for modeling the irregular and asynchronous nature of event streams, enabling graph neural networks to capture both local interactions and long-range dependencies.

\emph{\textbf{Point-based representations:}}
Point-based representations are the closest to the raw event stream. They represent events as a set of discrete spatio-temporal points:
\begin{equation}
    \mathcal{P} = \{\mathbf{q}_i\}_{i=1}^{N}, \quad
    \mathbf{q}_i = (x_i, y_i, t_i, p_i^{\ast}),
\end{equation}
where \(\mathbf{q}_i\) denotes the \(i\)-th event point, \(N\) denotes the number of events, \((x_i, y_i)\) and \(t_i\) denote its spatial location and timestamp, and \(p_i^{\ast}\) denotes an optional polarity attribute when polarity information is used.
As shown in Fig.~\ref{fig:representations}(f), point-based representations preserve the original spatio-temporal distribution of events without explicit frame conversion, making them suitable for event-native modeling.

\textbf{2) Multi-representation methods:}
Unlike single-representation methods that rely on one principal event encoding, multi-representation methods explicitly construct and jointly exploit multiple complementary event representations within the same perception pipeline.
Given an event stream \(\mathcal{E}\), a general multi-representation formulation can be expressed as:
\begin{equation}
    \mathcal{R}^{(m)} = \Phi_m(\mathcal{E}), \quad m=1,\ldots,M,
\end{equation}
where \(\Phi_m(\cdot)\) denotes the construction function of the \(m\)-th representation and \(M\geq2\).
The resulting representations are subsequently integrated through feature interaction, aggregation, or fusion mechanisms.

According to the structural types of the participating representations, existing methods are mainly organized into \textit{dense--dense} and \textit{dense--sparse} hybrid formulations.
Dense--dense methods combine multiple regular event encodings with complementary temporal or spatial properties, whereas dense--sparse methods jointly exploit the spatial regularity of dense representations and the fine-grained temporal or structural information retained by sparse representations.
Sparse--sparse combinations remain relatively underexplored and are therefore not treated as a separate category in this survey.

\textbf{Summary:}
Overall, this taxonomy characterizes event representation learning as a trade-off among information fidelity, structural regularity, computational efficiency, and architectural compatibility.

\section{Review Methodology}
\label{sec:methodology}

To provide a systematic overview of event camera representation learning, we conducted a structured literature search covering studies published up to \textit{August 2026}. Relevant papers were collected from IEEE Xplore, ScienceDirect, SpringerLink, arXiv, and Google Scholar using combinations of keywords including \textit{event camera}, \textit{event-based vision}, \textit{event representation}, \textit{event frame}, \textit{time surface}, \textit{voxel}, \textit{point cloud}, \textit{graph}, \textit{token}, and \textit{Transformer}, together with major downstream tasks such as classification, detection, tracking, segmentation, depth estimation, optical flow, and reconstruction. Citation tracing from influential surveys and representative studies was also used to identify additional relevant works.

The retrieved literature was screened according to its relevance to event representation construction, learning, optimization, or utilization. We included representation-oriented studies as the primary basis of the taxonomy, together with representative benchmark and application papers used to characterize task-level usage and empirical behavior. Duplicate or substantially overlapping versions, studies with insufficient methodological information, and works without a clearly identifiable event representation were excluded. After duplicate removal and title/abstract and full-text screening, more than 150 papers were retained and organized according to the taxonomy in Fig.~\ref{fig:taxonomy_overview_complex}, covering dense-based, sparse-based, and multi-representation paradigms.

\section{Differences from Existing Surveys}
In this section, we highlight the differences between our survey and existing reviews to clarify the significance of our work. Existing surveys on event-based vision mainly focus on sensor technology, general event-driven vision pipelines, specific application domains, or benchmark-oriented summaries of deep learning methods, while relatively few provide a dedicated and systematic review of \textbf{event camera representation learning}. The survey in \cite{chakravarthi2024recent} mainly emphasizes event camera hardware, sensing mechanisms, and emerging applications, with limited discussion of how raw event streams are transformed into learnable representations. The survey in \cite{gallego2020event} provides a broad overview of event-driven sensing, foundational principles, and representative tasks, but does not establish a representation-centric taxonomy. The survey in \cite{shariff2024event} focuses on automotive perception scenarios, while \cite{zheng2023deep} reviews deep learning frameworks, benchmark datasets, and evaluation settings, with greater emphasis on model performance and benchmark coverage than on the structural design space of event representations.

In contrast, our survey specifically focuses on how raw asynchronous event streams are transformed into learnable representations for modern neural architectures. We first distinguish methods according to whether they rely on a \textbf{single principal representation} or jointly exploit \textbf{multiple complementary representations}. Single-representation methods are further categorized into dense-based and sparse-based paradigms, while multi-representation methods are organized into dense--dense and dense--sparse hybrid formulations. Based on this taxonomy, we analyze their structural characteristics, design principles, modeling trade-offs, and task-level implications. We further summarize representative benchmarks and evaluation settings across high-level perception and low-level vision tasks, and discuss open challenges and future directions toward adaptive and efficient event representation learning. Overall, our survey provides a focused, representation-centric, and structurally unified perspective on event-based vision.

\section{Dense-based Single Representations}
Dense-based representations transform asynchronous event streams into regular grid-like structures, such as 2D maps, multi-channel tensors, or 3D spatio-temporal volumes.
Given an event stream \(\mathcal{E}=\{e_i\}_{i=1}^{N}\), where \(e_i=(x_i,y_i,t_i,p_i)\) denotes the \(i\)-th event, dense methods typically project irregular events onto structured arrays through temporal accumulation, timestamp overwriting, binning, or learned aggregation.
This conversion makes event data directly compatible with mature vision backbones, including CNNs, U-Nets, recurrent networks, and Transformers.
The price of this compatibility is that part of the microsecond-level temporal precision and native sparsity of Address-Event Representation (AER) is often traded for spatial continuity, batch processing efficiency, and architectural convenience.

\subsection{Event Frame}
\label{sec:ef}
\begin{figure}[t]
    \centering
    \includegraphics[width=\columnwidth]{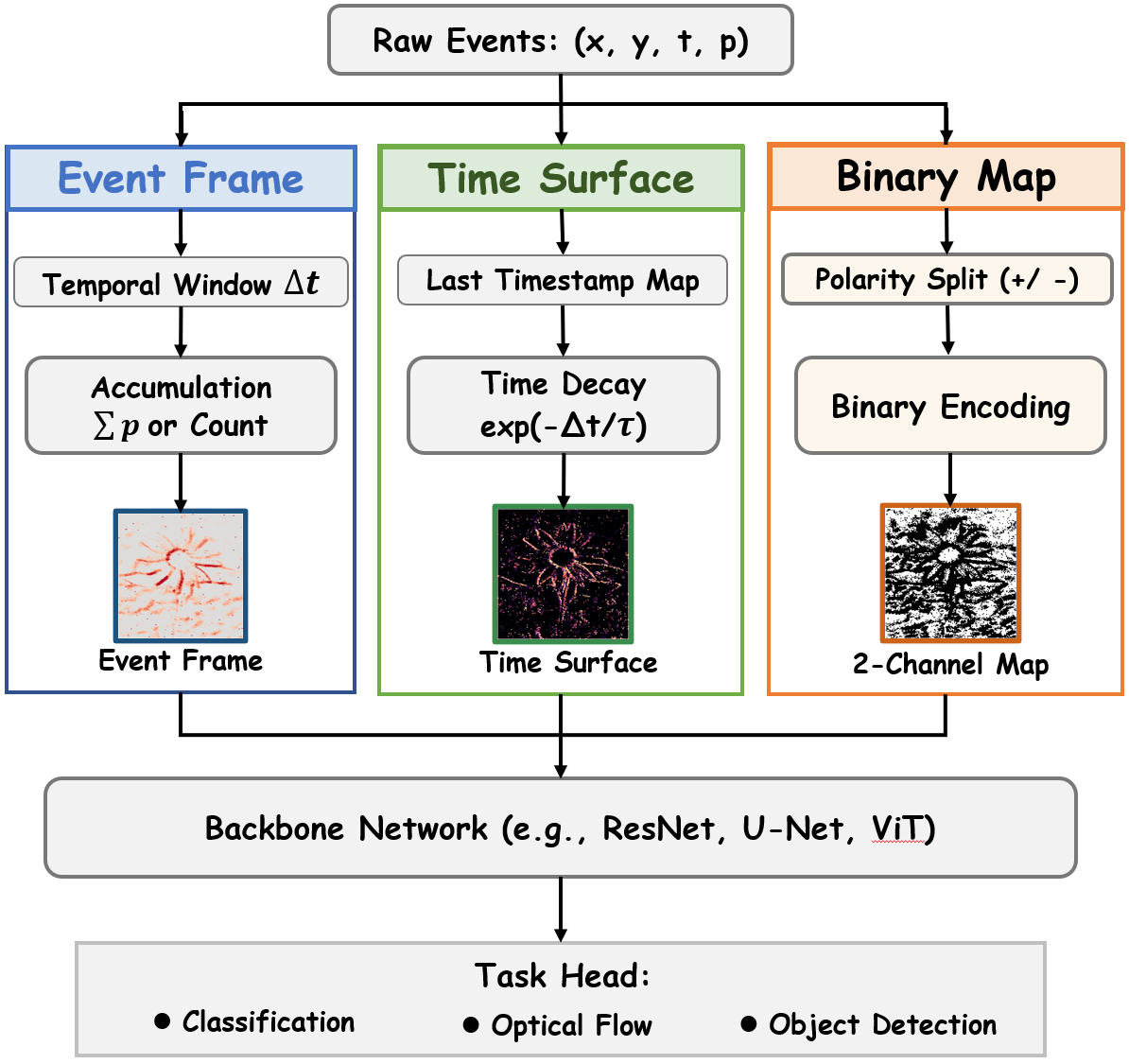}
    \caption{Overview of representative map-based representations. Raw events are projected into grid-structured formats, including Event Frame, Time Surface, and Binary Map, which emphasize event accumulation, temporal recency, and binary occupancy encoding, respectively.}
    \label{fig:frame_based_overview}
\end{figure}

The \textbf{Event Frame} is the most intuitive dense representation, treating the events occurring within a temporal interval as an image-like observation, as shown in Fig.\ref{fig:frame_based_overview}. 
It accumulates sparse events onto a two-dimensional spatial grid and therefore establishes a direct bridge between event cameras and conventional frame-based visual processing. Compared with raw event streams in AER form, event frames discard the exact ordering of events inside the integration window, but they provide a compact structure that can be processed by standard image-based vision networks.

Formally, given a temporal window $\Delta T=[t_s,t_e]$, an event frame can be constructed by accumulating events at their corresponding pixel locations. Since positive and negative events describe opposite brightness changes, most practical implementations adopt polarity-separated channels:
\begin{equation}
    H_{\pm}(x,y)=
    \sum_{e_i \in \mathcal{E},\, t_i \in \Delta T,\, p_i=\pm1}
    \delta(x-x_i,y-y_i),
    \label{eq:event_frame}
\end{equation}
where $\delta(\cdot)$ denotes the Kronecker delta function, and $H_{+}$ and $H_{-}$ record the numbers of positive and negative events, respectively. This formulation is closely related to early event histograms and pseudo-pictures~\cite{perez2013mapping,zhao2014feedforward}, which established the basic principle of converting asynchronous event streams into image-plane accumulations through polarity-aware counting.
Owing to their simplicity and architectural compatibility, event-frame representations have been widely adopted in representative event-based pipelines, including early frame-style mappings~\cite{perez2013mapping}, SNN-based recognition~\cite{amir2017low,shrestha2018slayer}, object tracking~\cite{wang2025mambaevt,wang2025event}, semantic segmentation~\cite{alonso2019evsegnet}, optical flow estimation~\cite{ding2022spatio}, and depth estimation~\cite{zhu2025depth}.

The representational behavior of event frames is mainly determined by the accumulation strategy. Fixed-time accumulation generates one frame within each constant temporal duration $\Delta T$~\cite{4444573,posch2010qvga,chen2011efficient}. This strategy is simple, deterministic, and naturally aligned with frame-based processing pipelines, but the resulting event density varies significantly with scene dynamics: static scenes may produce extremely sparse frames, whereas fast motion may lead to saturated contours and edge smearing. Fixed-event-number accumulation instead generates a frame after receiving a predefined number of events~\cite{delbruck2007fast,4444573}. By adapting the temporal length of the window to the event rate, this strategy produces more stable frame density under changing motion conditions, but it also makes the temporal duration of each frame scene-dependent.

Another important refinement is motion-aware accumulation. Instead of directly projecting all events in the window onto the same spatial plane, motion-compensated event frames align events to a reference timestamp before accumulation~\cite{rebecq2017real}. This design reveals that the limitation of event frames is not only caused by their two-dimensional grid structure, but also by whether the aggregation process respects the underlying motion. When the motion model is accurate, motion compensation can produce sharper spatial structures; when it is unavailable or inaccurate, temporally distinct observations are still collapsed into the same static map.

Overall, event frames provide a simple and efficient bridge between asynchronous events and image-based neural architectures. Their main advantage lies in low memory cost, easy implementation, and strong compatibility with existing vision backbones. However, their temporal aggregation discards fine-grained event ordering, making them less suitable for tasks that require precise temporal reasoning.

\subsection{Time Surface}
\label{sec:ts}
Different from event frames that emphasize event frequency, the \textbf{time surface} represents event streams through temporal recency. 
It records how recently each pixel has fired and converts asynchronous event streams into a two-dimensional map whose intensity reflects the freshness of local activity. This design preserves a stronger temporal cue than simple counting, because the value at each pixel is determined by the latest timestamp rather than the total number of accumulated events. 
Therefore, time surfaces can be viewed as recency-weighted dense memories of event streams.

The standard formulation follows the Surface of Active Events~\cite{benosman2013event}. For a pixel \((x,y)\), let \(t_{\mathrm{last}}(x,y)\) be the timestamp of the most recent event at that location before the current time \(t\). The time surface is defined as:
\begin{equation}
S(x,y,t)=\exp\left(-\frac{t-t_{\mathrm{last}}(x,y)}{\tau}\right),
\label{eq:time_surface_basic}
\end{equation}
where \(\tau\) is the decay constant controlling how quickly past activity fades. A small \(\tau\) emphasizes only very recent events, while a large \(\tau\) preserves longer motion traces.
In this sense, the key design variable of time surfaces is not event count, but the decay rule that determines how temporal history is retained and forgotten.

Several variants further refine the temporal memory mechanism of time surfaces. HOTS~\cite{lagorce2016hots} extends pixel-wise recency maps into a hierarchical representation, showing that local time-surface patterns can be progressively organized into higher-level spatio-temporal descriptors. Specifically, for an incoming event $e_i=(x_i,y_i,t_i,p_i)$, HOTS extracts a local time-surface patch around its spatial location:
\begin{equation}
    s_i(\Delta x,\Delta y)
    =
    \exp\left(
    -\frac{t_i-t_{\mathrm{last}}(x_i+\Delta x,y_i+\Delta y)}{\tau}
    \right),
    \label{eq:hots_patch_value}
\end{equation}
\begin{equation}
    \mathbf{s}_i =
    \{s_i(\Delta x,\Delta y)\mid(\Delta x,\Delta y)\in\Omega\}.
    \label{eq:hots_patch}
\end{equation}
where \(\Omega\) denotes the local neighborhood centered at \((x_i,y_i)\), \((\Delta x,\Delta y)\) indexes the relative spatial offsets within this neighborhood, and \(\mathbf{s}_i\) represents the recent spatio-temporal activity pattern around the current event.
HOTS learns hierarchical prototypes from such local time-surface patches, whereas HATS~\cite{sironi2018hats} aggregates them within spatial cells to obtain more stable local descriptors. Speed-invariant time surfaces~\cite{manderscheid2019speed} and Distance Surface~\cite{almatrafi2020distance} further reduce the dependence of recency values on motion speed by modifying the decay rule or replacing timestamp decay with distance to recent event structures. More recent adaptive and motion-aware designs~\cite{zhu2023event,xu2025mets,tang2026pa} adjust the decay process or encode motion cues to maintain stable temporal traces under varying event rates and spatial displacements. These developments suggest that the key challenge of time-surface representation is to preserve meaningful short-term motion memory while avoiding stale, speed-sensitive, or blurred structures.

Overall, time surfaces encode event streams through temporal recency rather than event counts, providing a compact description of short-term motion activity. They preserve more temporal information than simple event frames and are effective for local motion and feature patterns. However, because only the most recent timestamp is retained at each pixel, they remain limited in modeling long-range temporal dependencies.

\subsection{Binary Map}
\label{sec:bm}
The \textbf{Binary Map} representation encodes the presence or absence of events at each pixel within a temporal interval. 
Unlike event frames that accumulate event counts, binary maps focus on occupancy. A pixel is activated once at least one event occurs within the integration window.
A unified polarity-separated formulation can be written as:
\begin{equation}
b_k^p(x,y)=
\mathbb{I}\left[
\exists e_i\in\mathcal{E}:
(x_i,y_i)=(x,y),
t_i\in\tau_k,
p_i=p
\right],
\label{eq:binary_map}
\end{equation}
where \(\mathbb{I}[\cdot]\) denotes the indicator function, \(\tau_k\) denotes the \(k\)-th temporal sub-window of the full interval \(\Delta T\), \(k=1,\dots, K\), and \(p\in\{+1,-1\}\). When \(K=1\), this formulation reduces to a standard binary event map over the full interval \(\Delta T\); when \(K>1\), it produces an ordered sequence of binary occupancy maps.

Temporal Binary Representation (TBR)~\cite{9412991} introduced this bit-packing strategy by converting a sequence of binary event maps into a compact integer-valued representation:
\begin{equation}
B^p(x,y)=\sum_{k=1}^{K} b_k^p(x,y)\cdot 2^{k-1}.
\label{eq:tbr}
\end{equation}
Assuming \(\tau_1,\dots,\tau_K\) are ordered from early to late, different bits indicate whether a pixel is activated in different temporal sub-windows, allowing the representation to preserve coarse temporal order without storing a full temporal volume. Later, Bina-Rep~\cite{barchid2022bina} emphasized the effectiveness of simple binary event frames, showing that occupancy-based maps can provide a compact alternative to count-based accumulation. Together, these works indicate that binary event representations trade event counts and precise timestamps for efficient occupancy codes: TBR focuses on compact temporal ordering through bit packing, whereas Bina-Rep highlights the simplicity of binary activation maps.

Overall, binary maps provide compact occupancy-based event representations with low memory and computation costs. They are attractive for efficient recognition and hardware-friendly processing. However, their binary encoding discards event multiplicity, precise timing, and richer polarity interactions, limiting their representational capacity.

\subsection{Voxel Grid}
\label{sec:vg}

\begin{figure}[t]
    \centering
    \includegraphics[width=\columnwidth]{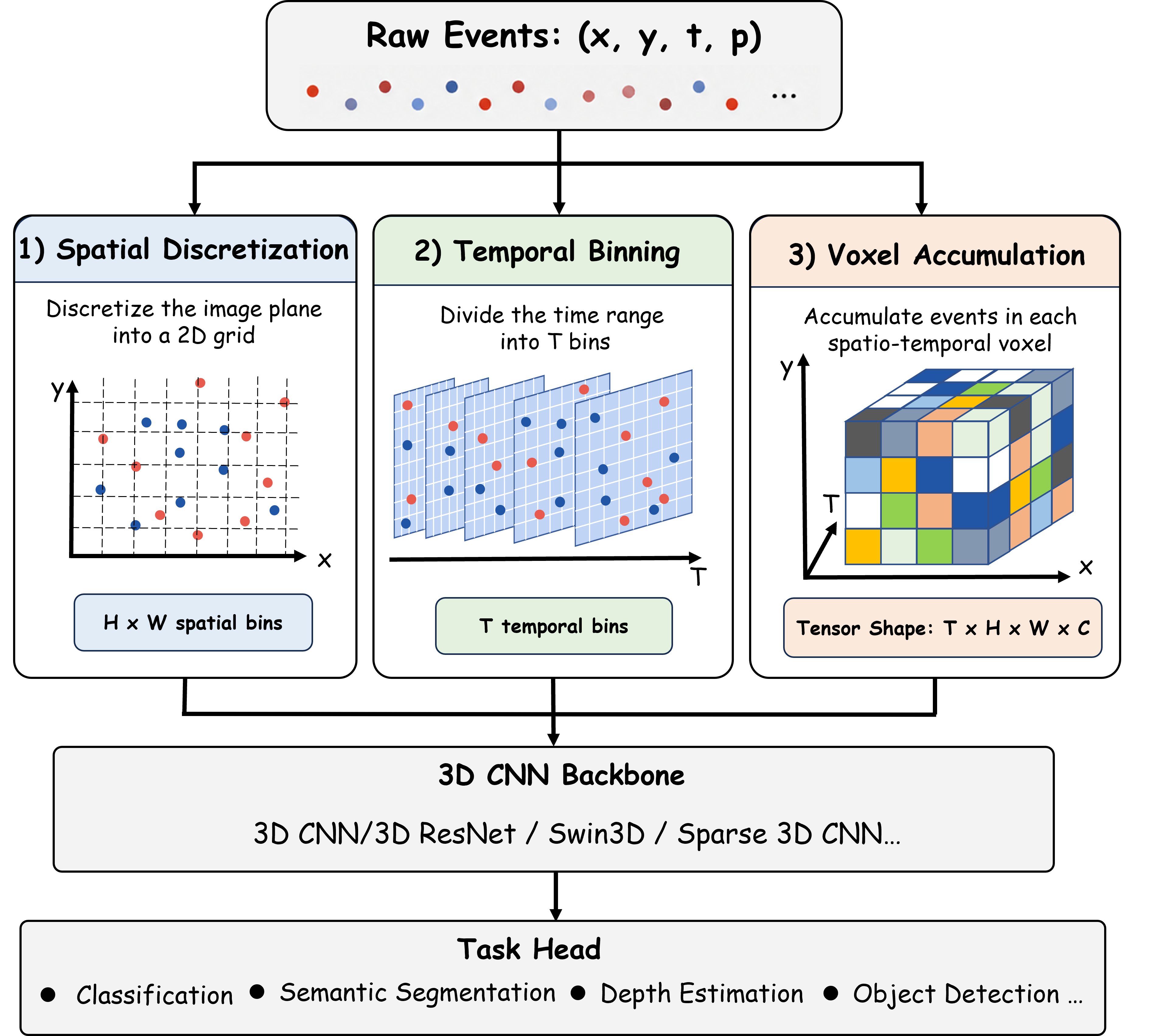}
    \caption{Overview of voxel-grid event representation. Raw events are discretized along spatial and temporal dimensions, accumulated into spatio-temporal voxels, and organized as a dense 3D tensor for downstream dense neural backbones and task heads.}
    \label{fig:voxel_based_overview}
\end{figure}
The \textbf{voxel-grid representation} is one of the most widely used dense spatio-temporal representations for event cameras. It organizes an event stream into a regular 3D volume, where the spatial axes correspond to image coordinates and the temporal axis is divided into multiple discrete bins. Unlike event frames that collapse all events within a time window into a single 2D map, voxel grids preserve a coarse temporal order by accumulating events into temporally ordered slices. By discretizing events along the temporal dimension, voxel grids transform asynchronous event streams into regular spatio-temporal tensors that can be processed by dense visual architectures, as shown in Fig.\ref{fig:voxel_based_overview}.

Formally, given an event window $\mathcal{E}_{[t_s,t_e]}$, where $t_s$ and $t_e$ denote the start and end timestamps of the event window, respectively, the timestamp of each event is first normalized to a continuous bin coordinate:
\begin{equation}
\tau_i=(B-1)\frac{t_i-t_s}{t_e-t_s},
\label{eq:voxel_time_norm}
\end{equation}
where $B$ denotes the number of temporal bins, and $\tau_i$ represents the normalized temporal position of event $e_i$ within the voxel grid. A standard voxel grid then accumulates events into a tensor $\mathbf{V}\in\mathbb{R}^{B\times H\times W}$ by:
\begin{equation}
\mathbf{V}_{b,x,y}
=
\sum_{e_i\in\mathcal{E}_{[t_s,t_e]}}
p_i\,\delta(x-x_i)\delta(y-y_i)\,\kappa(b-\tau_i),
\label{eq:voxel_grid}
\end{equation}
where $\mathbf{V}_{b,x,y}$ denotes the accumulated event value at temporal bin $b$ and spatial location $(x,y)$, $H$ and $W$ are the height and width of the sensor resolution, and $p_i$ is the polarity of event $e_i$. The spatial delta functions indicate that the event contributes only to its corresponding pixel location, while $\kappa(b-\tau_i)$ determines how the event is assigned to temporal bins according to its normalized timestamp. In practice, $\kappa(\cdot)$ can be implemented as a hard assignment kernel or a differentiable triangular interpolation kernel. Therefore, the construction of voxel grids is mainly determined by the temporal window size, the number of bins $B$, and the temporal assignment kernel $\kappa(\cdot)$.

The early development of voxel-grid representations was largely driven by dense motion estimation and event-based reconstruction. Differentiable voxel accumulation made it possible to connect asynchronous events with dense prediction networks~\cite{zhu2019unsupervised}, while event volumes further demonstrated their effectiveness for high-speed and high-dynamic-range video reconstruction~\cite{rebecq2019high}. These works explain why voxel grids became a default representation for low-level event vision: they retain coarse temporal variation useful for motion reasoning while preserving the regular spatial layout required by dense image restoration and prediction networks.

Subsequent studies extended voxel-grid representations beyond basic dense accumulation by combining them with stronger temporal modeling architectures. Multi-scale, Transformer-based, recurrent, sparse, and state-space designs further exploit the ordered voxel structure to capture spatio-temporal dependencies across bins and spatial locations~\cite{liu2023voxel,xie2024event,gehrig2023recurrent,peng2024scene,zubic2024state,yang2025smamba}. In this sense, voxel grids have evolved from a simple event-to-volume encoding into a general dense interface between event streams and modern neural architectures. Their regular image-plane alignment also makes them especially suitable for dense prediction and multimodal fusion, where outputs must remain spatially aligned with image coordinates~\cite{zhang2023multi,gehrig2021ramnet,zhang2023cmx,zhang2023delivering}.

Overall, voxel grids convert event streams into dense spatio-temporal tensors that preserve coarse temporal order while remaining compatible with mature neural architectures. Their advantage lies in structural compatibility: they retain spatial topology, support temporal modeling across bins, and can reuse dense backbones developed for image and video processing. However, temporal binning still loses microsecond-level ordering, memory cost increases with $B\times H\times W$, and fixed temporal windows may be suboptimal under highly variable event rates. Future voxel-grid research therefore requires more adaptive binning strategies, sparse or compressed voxel processing, and hybrid integration with token-, graph-, or point-based representations.
\subsection{Learned Dense Representation}
\label{sec:lr}
\begin{figure}[t]
    \centering
    \includegraphics[width=\columnwidth]{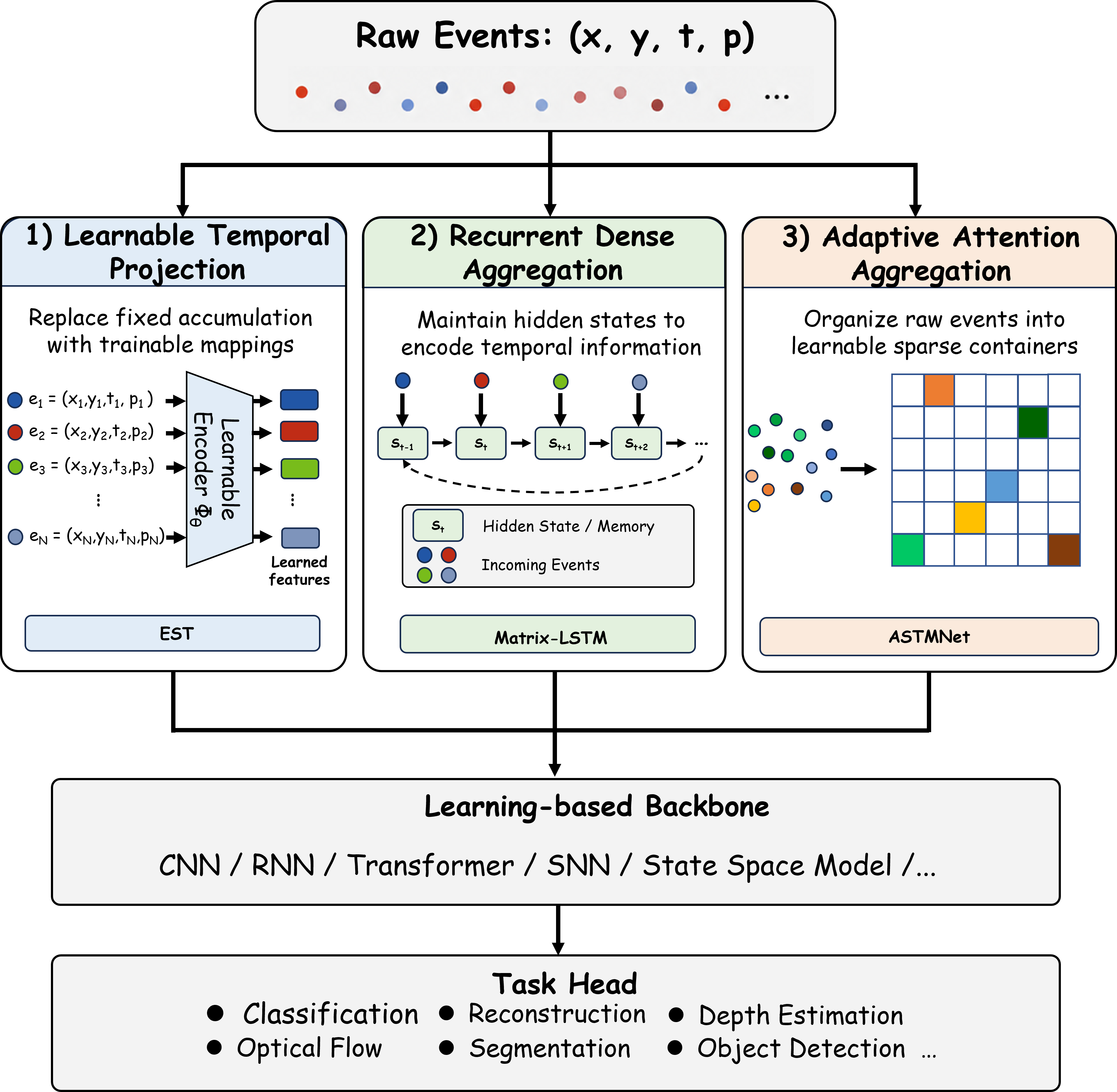}
    \caption{Illustration of learned dense event representation. Raw events are transformed into explicit dense representations through trainable event-to-grid mappings, including learnable temporal projection, recurrent aggregation, and adaptive attention-based integration, before being processed by downstream neural backbones.}
    \label{fig:learning_based_overview}
\end{figure}

The \textbf{learned dense representation} treats the event-to-grid mapping itself as a trainable component rather than a fixed preprocessing step.
Unlike handcrafted dense representations, such as event frames, time surfaces, binary maps, and voxel grids, whose construction relies on predefined accumulation, decay, thresholding, or temporal discretization rules, learned dense representations optimize how asynchronous events are projected and aggregated into explicit grid-structured representations.
This allows the representation construction process to adapt its temporal projection and feature aggregation according to data distribution and task supervision, as shown in Fig.\ref{fig:learning_based_overview}.

Formally, given an event stream
$\mathcal{E}=\{e_i\}_{i=1}^{N}$, where
$e_i=(x_i,y_i,t_i,p_i)$, a learned dense representation can be expressed as:
\begin{equation}
    \mathbf{Z}
    =
    \mathcal{R}_{\theta}(\mathcal{E}),
    \qquad
    \mathbf{Z}\in
    \mathbb{R}^{T\times H\times W\times C},
    \label{eq:learning_general}
\end{equation}
where $\mathcal{R}_{\theta}(\cdot)$ denotes a trainable event-to-representation mapping parameterized by $\theta$, and $\mathbf{Z}$ denotes the resulting dense representation.
Here, $T$, $H$, $W$, and $C$ denote the temporal dimension, height, width, and channel number, respectively.
In this survey, we restrict learned dense representations to methods in which the event-to-grid mapping itself is trainable and produces an explicit dense representation before the principal feature extractor.
Learned hidden states or task-network features alone are therefore not regarded as learned event representations.

Representative methods mainly differ in how the trainable event-to-grid mapping is constructed.
EST~\cite{gehrig2019end} replaces fixed temporal discretization with learnable basis functions, allowing event timestamps to contribute adaptively to dense feature channels.
Matrix-LSTM~\cite{cannici2020differentiable} further incorporates recurrent aggregation into representation construction, enabling asynchronous events to be accumulated into a learnable dense matrix representation while preserving temporal dependencies.
ASTMNet~\cite{li2022asynchronous} further introduces an asynchronous attention embedding that adaptively aggregates event streams into dense representations through learnable temporal sampling and attention-based integration.
Together, these methods illustrate the evolution from manually specified event accumulation toward increasingly adaptive and trainable dense representation construction.

Overall, learned dense representations retain the regular structure and architectural compatibility of dense event encodings while reducing reliance on manually designed aggregation rules.
Their flexibility, however, comes with increased data dependence and training complexity compared with fixed handcrafted representations.

\section{Sparse-based Single Representations}
The \textbf{sparse-based representation} aims to preserve the inherent sparsity of raw events by processing only informative event locations rather than constructing dense image-like structures. Unlike dense representations that populate regular grids regardless of event occupancy, sparse representations store and propagate features only at activated spatial or spatio-temporal positions, thereby reducing redundant computation and memory consumption. Depending on how sparsity is organized and exploited, existing approaches may represent events as irregular point sets, graph-structured entities, or compact token sequences. Although these formulations differ in their underlying data structures and neighborhood definitions, they share a common objective: maintaining the efficiency and fine-grained temporal characteristics of asynchronous events while enabling effective feature extraction and long-range dependency modeling.
\subsection{Token-based Representation}
\label{sec:tr}

\begin{figure}[t]
    \centering
    \includegraphics[width=0.95\columnwidth]{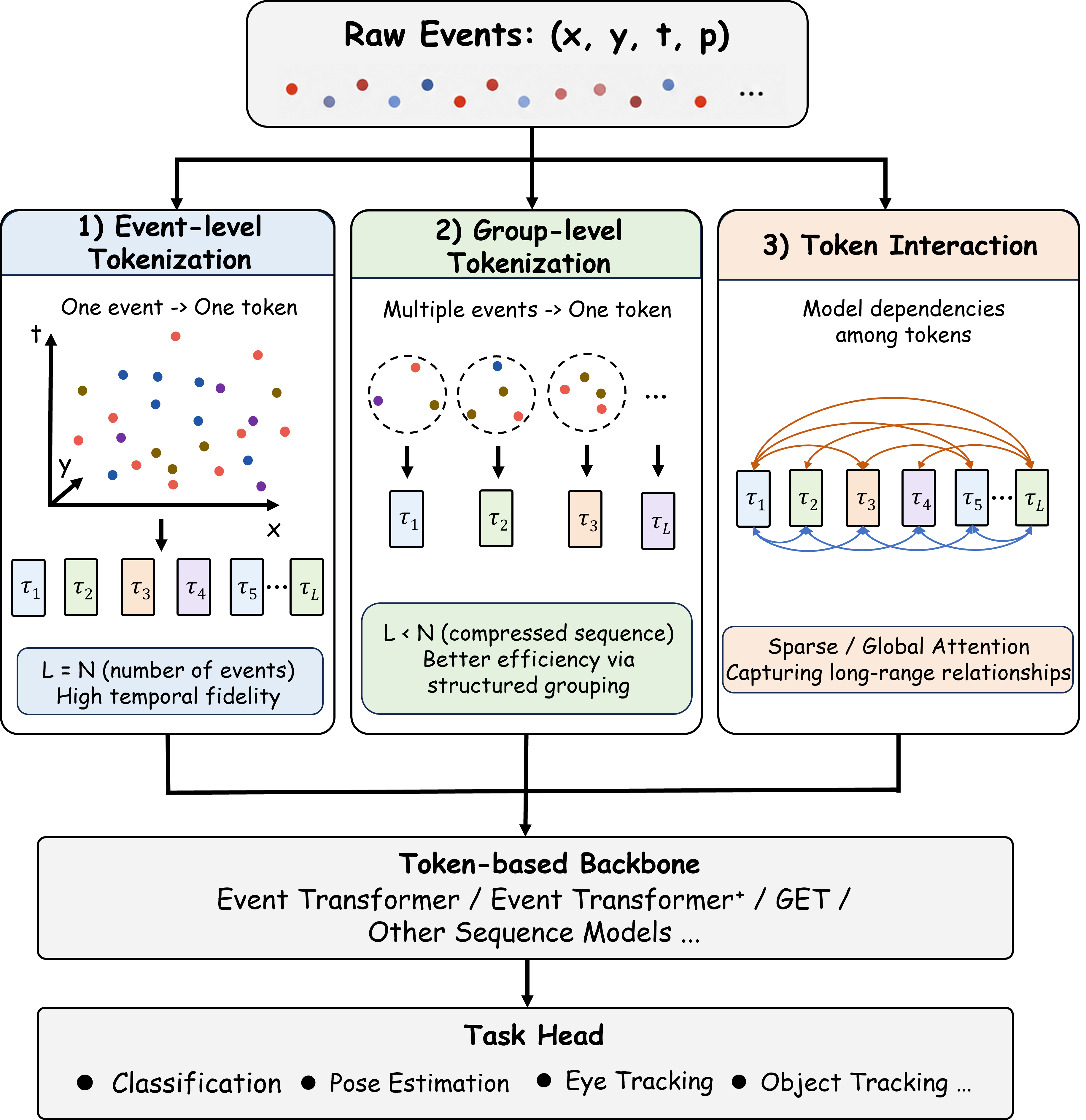}
    \caption{Illustration of token-based event representation. Raw events are organized into token sequences either by assigning each event to an individual token or by grouping multiple events into compact tokens, followed by token interaction modeling through sparse or global attention for downstream perception tasks.}
    \label{fig:token_based_overview}
\end{figure}

The \textbf{token-based representation} is a sequence-oriented event encoding paradigm that organizes raw event streams into data-dependent token sequences for sequence modeling. Unlike dense representations that rely on fixed grid structures, token-based representations abstract asynchronous events into variable-length token units directly from raw events without pre-generated event frames, voxel grids, or other handcrafted encodings. Depending on the tokenization strategy, each token may correspond to an individual event or a local event group. Tokenization, therefore, determines which event subsets are treated as atomic units for subsequent sequence interactions.

Formally, given an event stream $\mathcal{E}=\{e_i\}_{i=1}^{N}$, where each event is represented as $e_i=(x_i,y_i,t_i,p_i)$, a token-based representation constructs a token sequence:
\begin{equation}
\mathcal{T}
=
[\tau_1,\tau_2,\dots,\tau_L]
\in \mathbb{R}^{L\times D},
\label{eq}
\end{equation}
where $\mathcal{T}$ denotes the resulting token representation, $L$ is the number of tokens, and $D$ is the embedding dimension of each token. The value of $L$ depends on the tokenization strategy: event-level tokenization assigns one token to one event, whereas group-level tokenization aggregates multiple events into compact token units to improve efficiency. A general tokenization process can be expressed as:
\begin{equation}
\tau_k =
\mathcal{A}
(
\{
\operatorname{Emb}_{\theta}(e_i)
\mid
e_i\in\mathcal{G}_k
\}
)
+
\operatorname{PE}(\bar{x}_k,\bar{y}_k,\bar{t}_k,\bar{p}_k),
\label{eq}
\end{equation}
where $\mathcal{G}_k\subseteq\mathcal{E}$ denotes the subset of events assigned to the $k$-th token, $\operatorname{Emb}{\theta}(\cdot)$ maps raw event attributes into a latent feature space, $\mathcal{A}(\cdot)$ aggregates embedded event features within $\mathcal{G}_k$, and $\operatorname{PE}(\cdot)$ injects positional and polarity information derived from the associated event subset. This formulation reveals that token-based representations mainly differ in how event subsets are constructed, how local event information is aggregated, and how positional information is encoded. In this survey, we specifically restrict token-based representations to methods that \textbf{generate tokens directly from raw event streams} rather than from pre-generated dense representations, since tokenization in the latter primarily serves as a Transformer input format instead of a representation construction mechanism.

Early token-based studies primarily focused on enabling efficient attention over sparse event streams. Event Transformer~\cite{sabater2022event} partitions asynchronous events into local event sets and performs sparse-aware attention directly on the resulting event tokens, substantially reducing the computational burden of global attention while preserving long-range interactions. Event Transformer$^+$~\cite{sabater2023event} further generalizes this paradigm into a unified framework for event data processing, demonstrating that raw-event tokenization can support diverse event-based tasks within a common Transformer architecture.

Subsequent studies shifted the focus from efficient token interaction to token construction itself. GET~\cite{peng2023get} organizes asynchronous events into group tokens according to temporal and polarity cues, and employs Event Dual Self-Attention together with Group Token Aggregation to facilitate information exchange among complementary token groups. This design demonstrates that tokenization can serve as a task-aware compression mechanism rather than a passive format conversion step, as shown in Fig.~\ref {fig:token_based_overview}.

More recently, event-level tokenization has emerged to maximize temporal fidelity. Jiang et al.~\cite{jiang2024token} directly convert raw events into event tokens without temporal discretization and employ Transformer-based interactions to jointly model local and global dependencies. By preserving the original event granularity, this design maintains microsecond-level temporal cues at the cost of substantially longer token sequences.

Overall, token-based representations organize raw events or local event groups into compact embedding sequences for attention-based modeling. They provide a flexible way to capture long-range dependencies while reducing the need for dense grid construction. However, event-level tokens can be computationally expensive, and group-level tokens may lose fine-grained temporal details.

\subsection{Graph-based Representation}
\label{sec:gr}
\begin{figure}[t]
    \centering
    \includegraphics[width=\columnwidth]{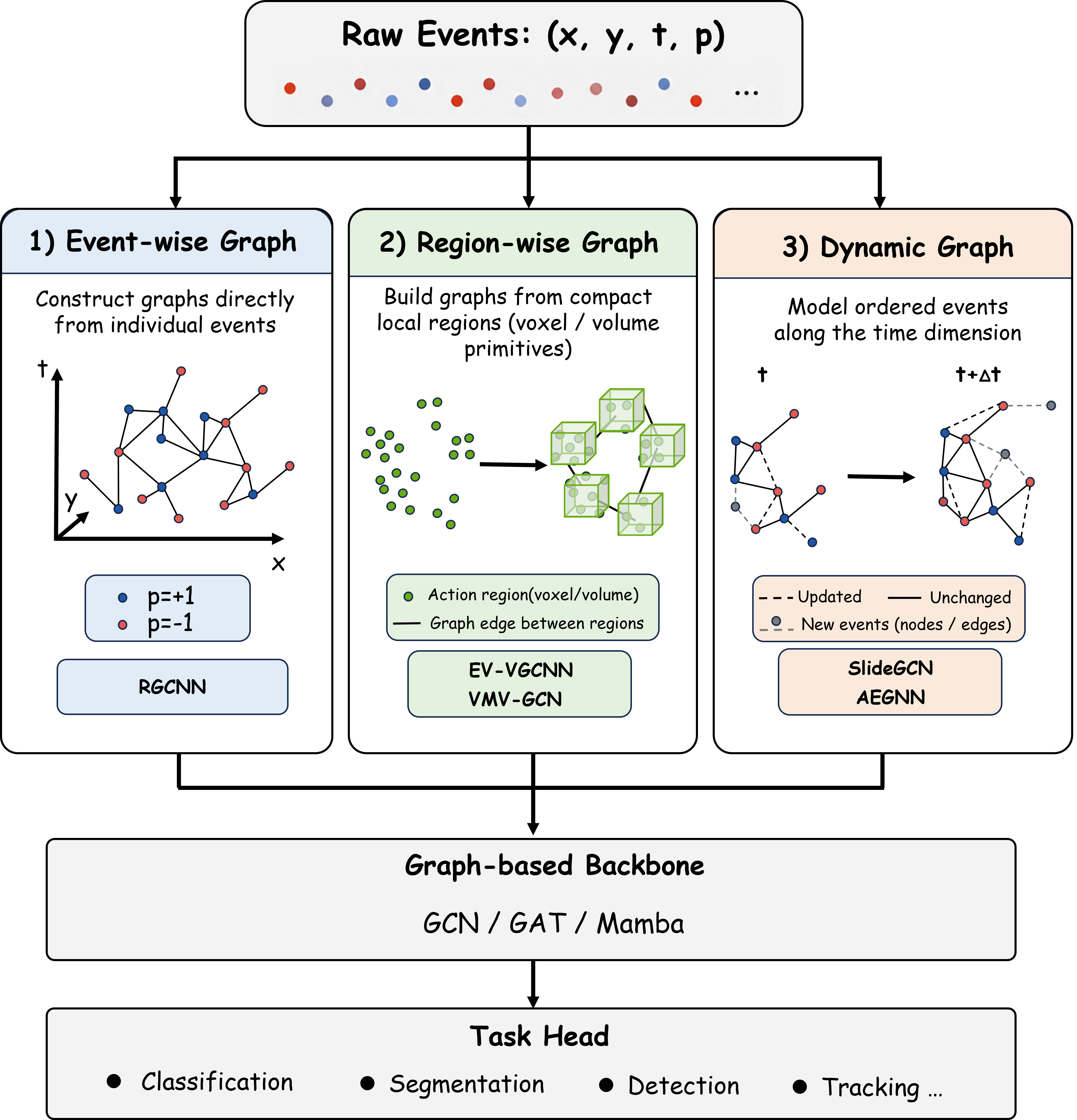}
    \caption{Illustration of graph-based event representation. Raw events are represented as graphs constructed either from individual events, compact local regions, or dynamically updated subgraphs, enabling graph-based backbones to capture spatial, temporal, and relational dependencies for downstream perception tasks.}
    \label{fig:graph_based_overview}
\end{figure}

The \textbf{graph-based representation} describes an event stream as a non-Euclidean relational structure. Since events are naturally sparse, asynchronous, and distributed along motion trajectories, representing them on regular grids may obscure their underlying topology and introduce redundant computation over empty regions. Instead of relying on fixed spatial layouts, graph-based representations explicitly model relational dependencies by constructing nodes and edges over sparse event structures. In this formulation, the central question is not only how to encode individual events, but also how to determine which events should exchange information.

Formally, given an event set
$\mathcal{E}=\{e_i=(x_i,y_i,t_i,p_i)\}_{i=1}^{N}$, a general graph-based representation can be written as:
\begin{equation}
    \mathcal{G}=(\mathcal{V},\mathcal{A},\mathbf{H}),\quad
    v_j=\mathcal{C}(\mathcal{S}_j),\quad
    h_j=\psi(v_j),
    \label{eq:graph_general}
\end{equation}
where $\mathcal{V}$ denotes the node set, $\mathcal{A}$ denotes the adjacency set, and $\mathbf{H}$ is the node feature matrix.
Here, $\mathcal{S}_j\subseteq\mathcal{E}$ denotes the event subset assigned to node $v_j$, $\mathcal{C}(\cdot)$ is the node construction operator, and $h_j$ is the node feature generated by $\psi(\cdot)$.
The adjacency set $\mathcal{A}$ is typically determined by spatial, temporal, spatio-temporal, or learned feature affinity, which controls how information is propagated among graph nodes.

Early graph-based representations explicitly construct relational structures over sparse event entities. 
In this event-wise graph formulation, nodes usually correspond to individual events, and edges are established according to spatial, temporal, or spatio-temporal proximity. 
RG-CNN~\cite{bi2020graph} constructs spatio-temporal graphs over event streams and aggregates neighborhood information through graph convolution, demonstrating that event relationships can be modeled through explicit message passing rather than dense spatial layouts. 
However, event-wise graph construction becomes computationally expensive under high event rates and is sensitive to isolated noisy events.

To improve scalability, subsequent studies construct graphs over compact local regions rather than individual events. 
In this region-wise formulation, graph nodes are derived from voxelized or volumetric event primitives, and only informative local regions participate in relational reasoning. 
EV-VGCNN~\cite{deng2022voxel} constructs graph vertices from voxelized event regions, while VMV-GCN~\cite{xie2022vmv} introduces volumetric multi-view graph modeling to capture complementary spatio-temporal structures. 
Although these methods may use voxelized primitives for node construction, they are still categorized as sparse graph representations because message passing is performed over selected nodes and edges rather than over a full dense lattice. 
This design reduces redundant computation while preserving structured neighborhood information.

A further line of work focuses on dynamic graph updating, which better matches the asynchronous nature of event streams. 
Instead of reconstructing the entire graph for each event window, SlideGCN~\cite{li2021graph} updates event relations through a sliding graph mechanism, while AEGNN~\cite{schaefer2022aegnn} updates only the local subgraph affected by newly arrived events. 
These methods show that graph-based representation depends not only on node and edge definitions, but also on how the relational structure evolves over time. 
More recently, EventMG~\cite{wu2025eventmg} further explores multilevel Mamba-Graph modeling, suggesting that dynamic graph representations may benefit from efficient sequence modeling for long-range dependency capture, as shown in Fig.~\ref {fig:graph_based_overview}.

Overall, graph-based representations model event streams as sparse relational structures rather than regular grids. They are effective for capturing local interactions and irregular spatio-temporal dependencies. However, their performance depends heavily on node construction, edge definition, and scalable graph updating under high event rates.

\subsection{Point-based Representation}
\label{sec:pr}

\begin{figure}[t]
    \centering
    \includegraphics[width=0.9\columnwidth]{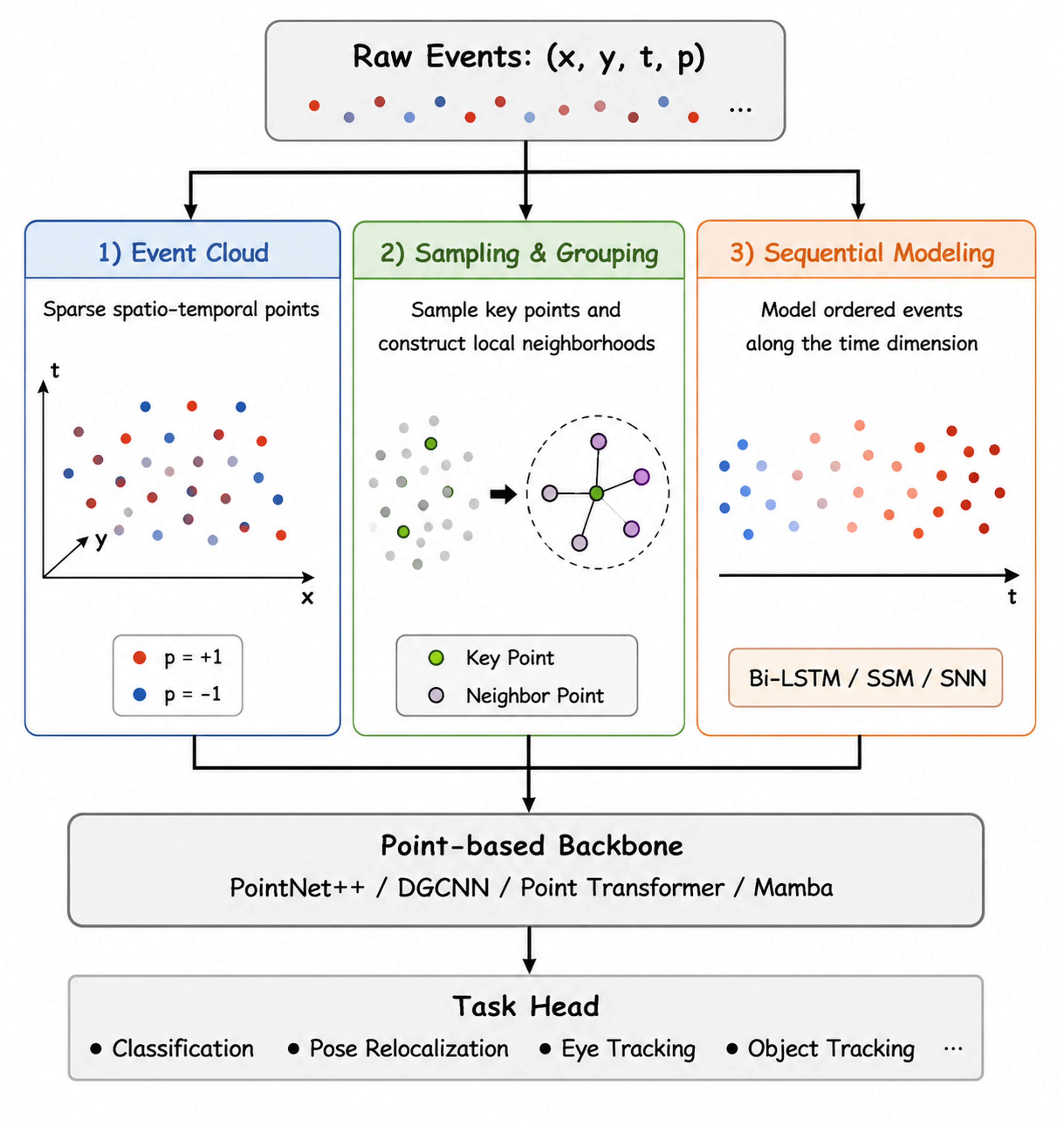}
    \caption{
    Overview of point-based event representations. 
    Point-based methods directly process sparse event clouds without converting events into dense grid structures. Typical pipelines include event cloud construction, local sampling and grouping, sequential temporal modeling, and point-based backbones for downstream event-based vision tasks.
    }
    \label{fig:point_based_overview}
\end{figure}

The \textbf{point-based representation} treats an event stream as an unordered set of sparse spatio-temporal points,
\[
\mathcal{P}=\{p_1,p_2,\dots,p_N\},
\]
where each point $p_i$ typically consists of spatial coordinates, timestamps, and polarity information. Unlike dense representations that require explicit projection onto regular grids, point-based representations operate directly on raw event streams without converting them into image-like or volume-like structures. By preserving the native sparsity and temporal precision of asynchronous events, they provide a natural framework for modeling fine-grained event dynamics directly from raw measurements, as shown in Fig.~\ref{fig:point_based_overview}.

A general point-based formulation can be written as:
\begin{equation}
\mathbf{F}
=
\gamma
\left(
\underset{p_i\in\mathcal{P}}{\mathcal{A}}
\left\{
h(p_i)
\right\}
\right),
\label{eq:point_agg}
\end{equation}
where $h(\cdot)$ denotes an event-wise feature embedding function, $\mathcal{A}(\cdot)$ represents a feature aggregation operation over the event set, and $\gamma(\cdot)$ denotes a task-specific mapping function. Different point-based representations mainly differ in how local neighborhoods are constructed, how temporal dependencies are modeled, and how aggregation is performed over large event sets.

Early point-based representations primarily focused on exploiting the geometric structure of event clouds. Inspired by point cloud learning frameworks such as PointNet~\cite{qi2017pointnet}, PointNet++~\cite{qi2017pointnet++}, and DGCNN~\cite{wang2019dynamic}, these methods regard events as sparse samples distributed in the spatio-temporal domain. ST-EC~\cite{wang2019space} introduced Space-time Event Clouds by treating timestamps as geometric coordinates and learning motion patterns directly from normalized event point distributions. PAT~\cite{yang2019modeling} further incorporates Gumbel Subset Sampling and self-attention to select representative event subsets while reducing computational complexity, enabling more effective modeling of long-range spatio-temporal interactions. EventNet~\cite{sekikawa2019eventnet} adopts recursive event-wise processing with temporal coding and hardware-friendly lookup tables, achieving real-time inference at event rates up to 1~MEPS. These studies establish the basic point-based paradigm in which event relationships are inferred through sparse neighborhood interactions rather than dense spatial layouts.

However, most early point-based methods primarily emphasize geometric feature extraction while underexploring the temporal dependencies embedded in asynchronous event streams. PEPNet~\cite{ren2024simple} combines hierarchical point-wise features with attentive Bi-LSTM aggregation to jointly model spatial and temporal cues by explicitly exploiting the sequential nature of event timestamps. EventMamba~\cite{ren2024rethinking} further integrates State Space Models into event cloud processing, enabling efficient modeling of long-range dependencies in chronologically ordered event sequences. These developments reflect an important transition from purely geometric reasoning toward explicit temporal modeling in point-based representations.

Recent studies increasingly focus on improving the efficiency and scalability of point-based representations. TTPOINT~\cite{ren2023ttpoint} compresses point feature extractors through Tensor Train Decomposition to achieve a favorable trade-off between recognition accuracy and computational efficiency. SpikePoint~\cite{ren2023spikepoint} combines point representations with spike-based computation to enable ultra-low-power event processing without frame conversion. E2B~\cite{ren2025e2b} demonstrates the effectiveness of raw event clouds for efficient object tracking through direct event-to-box prediction, while SECNet~\cite{ren2026scalable} introduces polarity-aware grouping together with Fourier-based feature extraction to improve scalability under high-resolution and long-duration event streams. These methods indicate a growing trend toward lightweight deployment and efficient processing of increasingly large event sets.

Overall, point-based representations preserve raw events as sparse spatio-temporal point sets without explicit frame or voxel conversion. They retain native sparsity and temporal precision, making them suitable for event-native modeling. However, they remain sensitive to sampling strategy, neighborhood construction cost, and limited spatial regularity.

\section{Multi-Representation Methods}
\subsection{Dense--Dense Hybrid}
Dense--dense hybrid representations jointly exploit multiple dense event encodings that preserve complementary aspects of the same event stream.
Although individual dense representations provide regular spatial organization and strong compatibility with conventional neural architectures, their behavior is highly dependent on how events are accumulated, temporally partitioned, or projected onto dense grids.
A single dense encoding may therefore emphasize only part of the available spatio-temporal information.
Dense--dense hybrid methods address this limitation by constructing multiple regular representations with different temporal granularities or aggregation characteristics and integrating them within a unified feature space.

Early studies demonstrated that complementary dense views can improve representation completeness by combining event observations formed under different accumulation strategies~\cite{gao2023action}.
EFV~\cite{yuan2023learning} further combines event-image and voxel representations, jointly exploiting compact spatial structure and temporally discretized event dynamics.
MTC~\cite{chen2019multi} combines complementary frequency-, recency-, and spike-inspired dense cues for event-based detection.
Subsequent methods extend this idea through representation-specific tensor construction and interaction mechanisms.
Spike-EVPR~\cite{liu2026spike} designs complementary dense tensors tailored to spiking processing, while EFV++~\cite{chen2025retain} and SMV-EAR~\cite{fan2026smv} further enhance information exchange across multiple dense views.
These developments indicate that dense--dense hybrid representation learning is gradually shifting from simple representation combination toward explicitly modeling complementary temporal and structural information.
However, constructing multiple dense representations also introduces additional memory, computation, and representation redundancy.

\subsection{Dense--Sparse Hybrid}

Dense--sparse hybrid representations combine the structural regularity of dense encodings with the fine-grained temporal and spatial information retained by sparse event structures.
Dense representations provide stable image-plane organization and efficient compatibility with mature neural backbones, whereas sparse representations preserve event-level distributions that may be weakened by temporal accumulation or discretization.
Their combination therefore provides a natural way to balance architectural compatibility and event-native representation fidelity.

Representative methods increasingly exploit this complementarity at different structural levels.
MTGA~\cite{zhang2025mtga} combines dense event frames with graph-structured event representations constructed at multiple temporal granularities, allowing regular spatial observations and sparse relational information to interact.
EventCrab~\cite{cao2024eventcrab} jointly models frame-based and point-based event representations, directly coupling dense spatial context with fine-grained event clouds.
More recently, MTGNet~\cite{lin2026event} integrates voxel and point representations across multiple temporal granularities, further showing that dense and sparse structures can provide complementary cues within the same event stream.
Overall, dense--sparse hybrid representations offer a promising compromise between structural regularity and temporal fidelity, but their effectiveness depends on how heterogeneous representations are aligned and interacted.
The additional construction and fusion overhead also motivates more adaptive representation selection and routing strategies, which will be discussed in the following section.
\begin{figure}[t]
    \centering
    \includegraphics[width=\columnwidth]{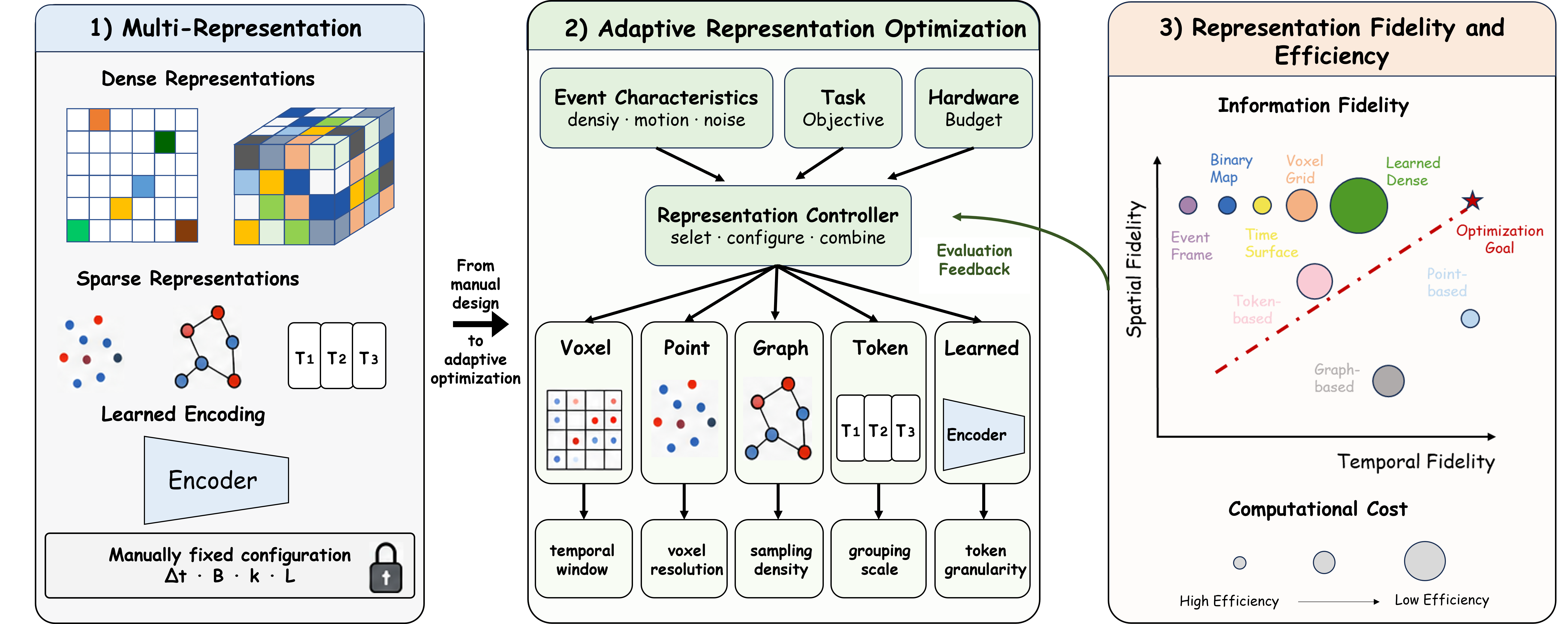}
\caption{From fixed multi-representation design to adaptive representation optimization, highlighting the trade-offs among temporal fidelity, spatial fidelity, and computational efficiency.}
    \label{fig:multi_based_overview}
\end{figure}

\section{From Representation Design to Representation Optimization}
\label{sec:rep_optimization}

As event representation families become increasingly mature, recent research is gradually shifting from designing new encoding formats toward optimizing how existing representations are constructed, selected, and evaluated. From this perspective, we highlight two closely related directions: adaptive representation selection and representation-centric analysis of fidelity and efficiency.

\subsection{Adaptive Representation Selection}

Most existing event-based systems adopt fixed representation configurations during training and inference. Once the temporal window, voxel resolution, point sampling strategy, graph neighborhood, or token granularity is determined, the same configuration is typically applied to all event streams. However, event density and temporal distribution vary substantially with motion, illumination, scene content, and sensor response, making a fixed configuration potentially inefficient or information-limited under changing conditions.

This motivates a shift from manually specified representation settings toward adaptive representation optimization. CHAOS~\cite{zubic2023chaos} shows that representation configurations can be systematically searched rather than selected purely by hand, providing an early step toward representation-level optimization. A natural extension is to make such decisions input-dependent, allowing temporal resolution, accumulation interval, sampling density, grouping scale, or token granularity to adapt according to the characteristics of the incoming event stream, as shown in Fig.\ref{fig:multi_based_overview} (b).

Adaptation may further extend across different representation forms. As reviewed in Section~VI, dense and sparse representations preserve complementary structural and temporal information, while EventDance~\cite{zheng2024eventdance} demonstrates the utility of cross-representation consistency. Future systems may therefore move from fixed representation choices or predetermined combinations toward adaptive routing, selection, and interaction among multiple representations. The key challenge is to obtain such flexibility without introducing excessive representation construction, conversion, or selection overhead.

\subsection{Representation Fidelity and Efficiency}

A second important direction concerns how event representations themselves should be evaluated. Existing studies predominantly rely on downstream metrics such as classification accuracy, detection mAP, optical-flow error, or reconstruction quality. Although these metrics provide useful empirical evidence, they are also affected by backbone capacity, training strategy, temporal windowing, supervision, and other implementation choices, making it difficult to isolate the contribution of the representation itself.

A more representation-centric evaluation should therefore consider both information fidelity and computational efficiency, as shown in Fig.\ref{fig:multi_based_overview} (c). Temporal fidelity reflects how much event timing and ordering is retained after accumulation, discretization, sampling, or compression, whereas spatial fidelity characterizes how well the original event-wise spatial distribution is preserved after discretization, sampling, grouping, or aggregation. Sparsity preservation further describes whether the event-driven structure of the input is maintained, while spatial regularity reflects the degree to which the resulting representation conforms to a structured spatial domain. Computational efficiency can be characterized by representation size, construction latency, memory consumption, and processing complexity.

These factors are inherently coupled. Dense representations preserve regular image-plane organization and spatial coverage, but often sacrifice fine-grained temporal information through accumulation or discretization. In contrast, point-, graph-, and token-based representations better retain event-level temporal information, while practical sampling, grouping, and structural abstraction may reduce spatial completeness and introduce irregular memory access or neighborhood construction overhead. Representation design should therefore be understood as a fidelity--efficiency trade-off rather than the pursuit of a universally optimal encoding. Future evaluation should compare representations under controlled computational settings to better distinguish gains introduced by representation construction from those brought by increasingly powerful backbones.

\section{Comparative Analysis of Empirical Results}
To connect the representation taxonomy with empirical evidence, Table~\ref{tab:comparison_all} summarizes representative results across different event-based vision tasks. These comparisons should not be interpreted as a strict leaderboard, since performance is affected not only by the representation itself, but also by backbone capacity, input modality, temporal windowing, training protocol, supervision, and dataset-specific settings. Nevertheless, several general observations can be drawn from the reported results.

First, representations that preserve stronger temporal fidelity, such as point-based, token-based, graph-based, or bit-level temporal encodings, often show advantages in scenarios where fine-grained motion dynamics and event ordering are important. Their strength comes from reducing the temporal collapse introduced by single-frame accumulation, although this usually increases the cost of sequence modeling, neighborhood construction, or attention computation. Second, dense representations, especially voxel-grid and related frame-like encodings, remain widely adopted when spatial alignment and pixel-level prediction are required. Their regular image-plane topology makes them compatible with mature convolutional, recurrent, and Transformer-based backbones, but this compatibility is obtained at the cost of temporal quantization and increased memory consumption. Third, compact representations such as event frames and binary maps remain valuable under latency, bandwidth, or hardware constraints, because they provide simple and efficient inputs even though they sacrifice event multiplicity or precise timestamps.

\section{Conclusion}
This survey reviews event camera representation learning from the perspective of how asynchronous event streams are transformed into learnable structures. We organize existing methods into single- and multi-representation paradigms, covering dense-based, sparse-based, and hybrid formulations. Our analysis highlights the trade-offs among temporal fidelity, structural regularity, sparsity preservation, computational efficiency, and architectural compatibility. Future research is expected to move toward more adaptive, scalable, and hardware-aware representation optimization for real-world event-based vision systems.

\clearpage
\onecolumn
\scriptsize
\setlength{\tabcolsep}{2.5pt}
\renewcommand{\arraystretch}{1.08}

\begin{longtable}{>{\raggedright\arraybackslash}p{0.10\textwidth}
                  >{\raggedright\arraybackslash}p{0.13\textwidth}
                  >{\raggedright\arraybackslash}p{0.27\textwidth}
                  >{\raggedright\arraybackslash}p{0.13\textwidth}
                  >{\raggedright\arraybackslash}p{0.15\textwidth}
                  >{\raggedright\arraybackslash}p{0.14\textwidth}}
\caption{Comprehensive comparison of event-based vision tasks. Methods are grouped by representation type within each dataset, and the best overall result within each dataset is highlighted.}
\label{tab:comparison_all} \\
\toprule
\textbf{Specific Task} & \textbf{Dataset} & \textbf{Method} & \textbf{Rep. Type} & \textbf{Metric} & \textbf{Performance} \\
\midrule
\endfirsthead

\multicolumn{6}{c}{\textit{Table \thetable\ continued}} \\
\toprule
\textbf{Specific Task} & \textbf{Dataset} & \textbf{Method} & \textbf{Rep. Type} & \textbf{Metric} & \textbf{Performance} \\
\midrule
\endhead

\midrule
\multicolumn{6}{r}{\textit{Continued on next page}} \\
\endfoot

\bottomrule
\endlastfoot

\multirow{40}{*}{Action Recognition} 
    & \multirow{17}{*}{DVS128 Gesture\cite{amir2017low}} 
      & Deep SNN \cite{amir2017low} & Event Frame & Acc. (\%) $\uparrow$ & 91.80 \\
    & & SLAYER \cite{shrestha2018slayer} & Event Frame & Acc. (\%) $\uparrow$ & 93.64 \\
    & & TA-SNN \cite{yao2021temporal} & Event Frame & Acc. (\%) $\uparrow$ & 98.61 \\
    & & STSC-SNN \cite{xu2023stsc} & Event Frame & Acc. (\%) $\uparrow$ & 98.96 \\
    & & TCJA-SNN \cite{zhu2022tcja} & Event Frame & Acc. (\%) $\uparrow$ & 99.00 \\
    & & TBR \cite{9412991} & Binary Map & Acc. (\%) $\uparrow$ & \textbf{99.58} \\
    & & VMST-Net \cite{liu2023voxel} & Voxel Grid & Acc. (\%) $\uparrow$ & 97.80 \\
    & & EVSTr \cite{xie2024event} & Voxel Grid & Acc. (\%) $\uparrow$ & 98.60 \\
    & & ACE-BET \cite{liu2022fast} & Voxel Grid & Acc. (\%) $\uparrow$ & 98.88 \\
    & & Motion-SNN \cite{liu2021eventaction} & Event Frame & Acc. (\%) $\uparrow$ & 92.70 \\
    & & GET \cite{peng2023get} & Token-based & Acc. (\%) $\uparrow$ & 97.90 \\
    & & EventTransformer \cite{sabater2022event} & Token-based & Acc. (\%) $\uparrow$ & 96.20 \\
    & & RG-CNN \cite{bi2020graph} & Graph-based & Acc. (\%) $\uparrow$ & 97.20 \\
    & & EDGCN+CRD \cite{deng2024dynamic} & Graph-based & Acc. (\%) $\uparrow$ & 98.30 \\
    & & ST-EC \cite{wang2019space} & Point-based & Acc. (\%) $\uparrow$ & 97.08 \\
    & & SpikePoint \cite{ren2023spikepoint} & Point-based & Acc. (\%) $\uparrow$ & 98.74 \\
    & & TTPOINT \cite{ren2023ttpoint} & Point-based & Acc. (\%) $\uparrow$ & 98.80 \\
    & & SECNet \cite{ren2026scalable} & Point-based & Acc. (\%) $\uparrow$ & 98.90 \\
    & & EventMamba \cite{ren2024rethinking} & Point-based & Acc. (\%) $\uparrow$ & 99.20 \\

\cmidrule{2-6}
 
    & \multirow{9}{*}{DailyAction-DVS\cite{liu2021eventaction}} 
    & HMAX-SNN \cite{xiao2019hmax} & Time Surface & Acc. (\%) $\uparrow$ & 68.30 \\
    & & HMAX-SPA \cite{liu2020event} & Time Surface & Acc. (\%) $\uparrow$ & 76.90 \\
    & & EVSTr \cite{xie2024event} & Voxel Grid & Acc. (\%) $\uparrow$ & 99.60 \\
    & & Motion-SNN \cite{liu2021eventaction} & Event Frame & Acc. (\%) $\uparrow$ & 90.30 \\
    & & VMV-GCN \cite{xie2022vmv} & Graph-based & Acc. (\%) $\uparrow$ & 94.10 \\
    & & SpikePoint \cite{ren2023spikepoint} & Point-based & Acc. (\%) $\uparrow$ & 97.92 \\
    & & EventMamba \cite{ren2024rethinking} & Point-based & Acc. (\%) $\uparrow$ & 99.10 \\
    & & TTPOINT \cite{ren2023ttpoint} & Point-based & Acc. (\%) $\uparrow$ & 99.10 \\
    & & SECNet \cite{ren2026scalable} & Point-based & Acc. (\%) $\uparrow$ & \textbf{99.65} \\

\cmidrule{2-6}
 
    & \multirow{5}{*}{DailyDVS-200\cite{wang2024dailydvs}} 
      
     & EST \cite{gehrig2019end} & Learned Dense & Top-1 Acc. (\%) $\uparrow$ & 32.23 \\
    & & ESTF \cite{gao2024hardvs} & Token-based & Top-1 Acc. (\%) $\uparrow$ & 24.68 \\
    & & GET \cite{peng2023get} & Token-based & Top-1 Acc. (\%) $\uparrow$ & 37.28\\
    & & SMV-EAR \cite{fan2026smv} & Dense–Dense Hybrid & Top-1 Acc. (\%) $\uparrow$ & \textbf{54.65} \\

    
    \midrule
    \multirow{26}{*}{Classification} 
    & \multirow{11}{*}{N-MNIST\cite{orchard2015converting}} 
     & SNN-BP \cite{lee2016training} & Event Frame & Acc. (\%) $\uparrow$ & 98.74 \\
    & & EFV \cite{yuan2023learning} & Dense--Dense Hybrid & Acc. (\%) $\uparrow$ & 98.90 \\
    & & RG-CNN \cite{bi2020graph} & Graph-based & Acc. (\%) $\uparrow$ & 99.00 \\
    & & HATS \cite{sironi2018hats} & Time Surface & Acc. (\%) $\uparrow$ & 99.10 \\
    & & ECSNet \cite{chen2022ecsnet} & Time Surface & Acc. (\%) $\uparrow$ & 99.20 \\
    & & Bina-Rep \cite{barchid2022bina} & Binary Map & Acc. (\%) $\uparrow$ & 99.34 \\
    & & VMST-Net \cite{liu2023voxel} & Voxel Grid & Acc. (\%) $\uparrow$ & 99.50 \\
    & & CTN \cite{zhao2022transformer} & Event Frame & Acc. (\%) $\uparrow$ & 99.70 \\
    & & GET \cite{peng2023get} & Token-based & Acc. (\%) $\uparrow$ & 99.70 \\
    & & SECNet \cite{ren2026scalable} & Point-based & Acc. (\%) $\uparrow$ & 99.70 \\
    & & ETB \cite{jiang2024token} & Token-based & Acc. (\%) $\uparrow$ & \textbf{99.90} \\
    \cmidrule{2-6}
    
    & \multirow{12}{*}{N-Caltech101\cite{orchard2015converting}} 
      & HATS \cite{sironi2018hats} & Time Surface & Acc. (\%) $\uparrow$ & 64.20 \\
    & & EVSTr \cite{xie2024event} & Voxel Grid & Acc. (\%) $\uparrow$ & 79.70 \\
    & & CES \cite{lin2025compressed} & Voxel Grid & Acc. (\%) $\uparrow$ & 84.70 \\
    & & E2VID \cite{rebecq2019high} & Voxel Grid & Acc. (\%) $\uparrow$ & 86.60 \\
    & & ACE-BET \cite{liu2022fast} & Voxel Grid & Acc. (\%) $\uparrow$ & 89.95 \\
    & & EST \cite{gehrig2019end} & Learned Dense & Acc. (\%) $\uparrow$ & 81.70 \\
    & & Matrix-LSTM \cite{cannici2020differentiable} & Learned Dense & Acc. (\%) $\uparrow$ & 85.72 \\
    & & ETB \cite{jiang2024token} & Token-based & Acc. (\%) $\uparrow$ & 81.60 \\
    & & RG-CNN \cite{bi2020graph} & Graph-based & Acc. (\%) $\uparrow$ & 65.70 \\
    & & EFV++ \cite{chen2025retain} & Dense–Dense Hybrid & Acc. (\%) $\uparrow$ & \textbf{92.1} \\
    & & EV-VGCNN \cite{deng2022voxel} & Graph-based & Acc. (\%) $\uparrow$ & 74.80 \\
    & & AEGNN \cite{schaefer2022aegnn} & Graph-based & Acc. (\%) $\uparrow$ & 66.80 \\
    & & SlideGCN \cite{li2021graph} & Graph-based & Acc. (\%) $\uparrow$ & 76.10 \\
    & & EDGCN+CRD \cite{deng2024dynamic} & Graph-based & Acc. (\%) $\uparrow$ & 83.50 \\
    & & SECNet \cite{ren2026scalable} & Point-based & Acc. (\%) $\uparrow$ & 82.40 \\

\midrule
\multirow{20}{*}{Object Tracking} 
    & \multirow{5}{*}{FE240hz} 
     & MambaEVT \cite{wang2025mambaevt} & Event Frame & SR/PR $\uparrow$ & 58.09/91.97 \\
    & & HDETrack \cite{wang2024event} & Event Frame & SR/PR $\uparrow$ & 59.80/92.20 \\
    & & HDETrack V2 \cite{wang2025event} & Event Frame & SR/PR $\uparrow$ & \textbf{60.40/93.20} \\
    & & STNet \cite{zhang2022spiking} & Event Frame & SR/PR $\uparrow$ & 58.50/89.60 \\
    & & SFTrack-Slow \cite{wang2025towards} & Graph-based & SR/PR $\uparrow$ & 59.90/92.70 \\

\cmidrule(lr){2-6}
 
    & \multirow{8}{*}{VisEvent \cite{wang2023visevent}} 

     & MambaEVT \cite{wang2025mambaevt} & Event Frame & SR/PR/NPR $\uparrow$ & 35.90/50.20/39.40 \\
    & & MambaEVT-P \cite{wang2025mambaevt} & Event Frame & SR/PR/NPR $\uparrow$ & 37.20/51.80/40.80 \\
    & & HDETrack \cite{wang2024event} & Event Frame & SR/PR/NPR $\uparrow$ & 37.30/54.60/44.50 \\
    & & HDETrack V2 \cite{wang2025event} & Event Frame & SR/PR/NPR $\uparrow$ & 38.00/55.80/45.10 \\
    & & CRSOT \cite{zhu2025crsot} & Event Frame & SR/PR $\uparrow$ & 52.50/74.10 \\
    & & CMT-MDNet (MF) \cite{wang2023visevent} & Event Frame & SR/PR/NPR $\uparrow$ & 57.44/67.20/69.78 \\
    & & CEUTrack \cite{tang2025revisiting} & Voxel Grid & SR/PR/NPR $\uparrow$ & \textbf{64.89/69.06/73.81} \\
    & & STNet (Event-Only) \cite{zhang2022spiking} & Event Frame & SR/PR/NPR $\uparrow$ & 39.70/49.20/-- \\

\cmidrule(lr){2-6}
    & \multirow{5}{*}{COESOT \cite{tang2025revisiting}} 
    & HDETrack \cite{wang2024event} & Event Frame & SR/PR/NPR $\uparrow$ & 53.10/64.10/64.50 \\
    & & CMT \cite{wang2023visevent} & Event Frame & SR/PR $\uparrow$ & \textbf{53.30/66.50} \\
    & & MDNet \cite{wang2023visevent} & Event Frame & SR/PR $\uparrow$ & \textbf{53.30/66.50} \\
    & & SFTrack-Fast \cite{wang2025towards} & Graph-based & SR/PR/NPR $\uparrow$ & 49.30/59.10/59.80 \\
    & & SFTrack-Slow \cite{wang2025towards} & Graph-based & SR/PR/NPR $\uparrow$ & 51.80/62.90/62.90 \\

\cmidrule(lr){2-6}
 
    & \multirow{2}{*}{EventVOT} 
      & HDETrack (Base) \cite{wang2024event} & Event Frame & SR/PR/NPR $\uparrow$ & 55.40/60.40/71.10 \\
    & & HDETrack \cite{wang2024event} & Event Frame & SR/PR/NPR $\uparrow$ & \textbf{57.80/62.20/73.50} \\

\cmidrule(lr){2-6}
    & \multirow{5}{*}{FE108 \cite{zhang2021object}} 
    & CMT-ATOM \cite{wang2023visevent} & Event Frame & SR/PR $\uparrow$ & 54.30/79.40 \\
    & & Mamba-FETrack \cite{huang2024mamba} & Event Frame & SR/PR $\uparrow$ & 58.71/90.95 \\
    & & Mamba-FETrack V2 \cite{wang2025mamba} & Event Frame & SR/PR $\uparrow$ & \textbf{62.18/94.74} \\
    & & CEUTrack \cite{tang2025revisiting} & Voxel Grid & SR/PR $\uparrow$ & 55.58/84.46 \\
    & & Zhu et al. \cite{zhu2022learning} & Graph-based & SR/PR $\uparrow$ & 54.90/85.90 \\

 

\midrule
\multirow{67}{*}{\makecell[l]{Automotive\\Detection}} 
    & \multirow{8}{*}{\makecell[l]{PKU-DDD17\\-Car\cite{binas2017ddd17,cao2024embracing}}} 
    & RAMNet-fusion \cite{cao2024embracing} & Event Frame & mAP50 (\%) $\uparrow$ & 79.6 \\
    & & JDF \cite{li2019event} & Event Frame & mAP50 (\%) $\uparrow$ & 84.1 \\
    & & CAFR \cite{cao2024embracing} & Event Frame & mAP50 (\%) $\uparrow$ & \textbf{86.7} \\
    & & MTC \cite{chen2019multi} & Dense–Dense Hybrid & mAP50 (\%) $\uparrow$ & 47.8 \\
    & & EFNet \cite{cao2024embracing} & Voxel Grid & mAP50 (\%) $\uparrow$ & 83.0 \\
    & & SPNet-fusion \cite{cao2024embracing} & Voxel Grid & mAP50 (\%) $\uparrow$ & 84.7 \\
    & & FAGC \cite{cao2024embracing} & Voxel Grid & mAP50 (\%) $\uparrow$ & 84.8 \\
    & & ASTMNet \cite{li2022asynchronous} & Learned Dense & mAP50 (\%) $\uparrow$ & 46.2 \\

\cmidrule{2-6}
 
    & \multirow{23}{*}{1Mpx\cite{perot2020learning}} 
      & SAM \cite{liang2021event} & Event Frame & mAP (\%) $\uparrow$ & 23.9 \\
    & & AEC \cite{peng2023better} & Voxel Grid & mAP (\%) $\uparrow$ & 48.4 \\
    & & MAD-Det \cite{chen2025motion} & Dense--Dense Hybrid & mAP (\%) $\uparrow$ & \textbf{49.5} \\
    & & EventPillars \cite{fan2025eventpillars} & Voxel Grid & mAP (\%) $\uparrow$ & 44.4 \\
    & & LEOD-RVT-S \cite{wu2024leod} & Voxel Grid & mAP (\%) $\uparrow$ & 46.7 \\
    & & SAST \cite{peng2024scene} & Voxel Grid & mAP (\%) $\uparrow$ & 48.3 \\
    & & SAST-CB \cite{peng2024scene} & Voxel Grid & mAP (\%) $\uparrow$ & 48.7 \\
    & & SMamba \cite{yang2025smamba} & Voxel Grid & mAP (\%) $\uparrow$ & 49.3 \\
    & & ERGO-12 \cite{zubic2023chaos} & Dense(Optimized) & mAP (\%) $\uparrow$ & 40.6 \\
    & & RED \cite{perot2020learning} & Voxel Grid & mAP (\%) $\uparrow$ & 43.0 \\
    & & ASTMNet \cite{li2022asynchronous} & Learned Dense & mAP (\%) $\uparrow$ & 48.3 \\
    & & GET-T \cite{peng2023get} & Token-based & mAP (\%) $\uparrow$ & 48.4 \\
    & & EventMG \cite{wu2025eventmg} & Graph-based & mAP (\%) $\uparrow$ & 47.4 \\

\cmidrule{2-6}
 
    & \multirow{19}{*}{Gen1\cite{de2020large}} 
      & AsyNet \cite{messikommer2020event} & Event Frame & mAP (\%) $\uparrow$ & 14.5 \\
    & & SAM \cite{liang2021event} & Event Frame & mAP (\%) $\uparrow$ & 35.5 \\
    & & AEC \cite{peng2023better} & Voxel Grid & mAP (\%) $\uparrow$ & 47.0 \\
    & & MAD-Det \cite{chen2025motion} & Dense--Dense Hybrid & mAP (\%) $\uparrow$ & 49.2 \\
    & & Spiking DenseNet \cite{cordone2022object} & Voxel Grid & mAP (\%) $\uparrow$ & 18.9 \\
    & & RVT-B \cite{gehrig2023recurrent} & Voxel Grid & mAP (\%) $\uparrow$ & 47.2 \\
    & & SpikeFPN \cite{zhang2024automotive} & Voxel Grid & mAP50 (\%) $\uparrow$ & 47.7 \\
    & & SAST-CB \cite{peng2024scene} & Voxel Grid & mAP (\%) $\uparrow$ & 48.2 \\
    & & LEOD-RVT-S \cite{wu2024leod} & Voxel Grid & mAP (\%) $\uparrow$ & 48.7 \\
    & & SMamba \cite{yang2025smamba} & Voxel Grid & mAP (\%) $\uparrow$ & 50.4 \\
    & & EventPillars \cite{fan2025eventpillars} & Voxel Grid & mAP (\%) $\uparrow$ & 53.1 \\
    & & Matrix-LSTM \cite{cannici2020differentiable} & Learned Dense & mAP (\%) $\uparrow$ & 31.0 \\
    & & RED \cite{perot2020learning} & Voxel Grid & mAP (\%) $\uparrow$ & 40.0 \\
    & & ASTMNet \cite{li2022asynchronous} & Learned Dense & mAP (\%) $\uparrow$ & 46.7 \\
    & & ERGO-12 \cite{zubic2023chaos} & Dense(Optimized) & mAP (\%) $\uparrow$ & 50.4 \\
    & & SSLA-Det \cite{hao2026ssla} & Token-based & mAP (\%) $\uparrow$ & 37.5 \\
    & & GET-T \cite{peng2023get} & Token-based & mAP (\%) $\uparrow$ & 47.9 \\
    & & AEGNN \cite{schaefer2022aegnn} & Graph-based & mAP (\%) $\uparrow$ & 16.3 \\
    & & EventMG \cite{wu2025eventmg} & Graph-based & mAP (\%) $\uparrow$ & \textbf{53.7} \\

\midrule
\multirow{13}{*}{\makecell[l]{Semantic\\Segmentation}} 
    & \multirow{4}{*}{EV-IMO\cite{8968520}} 
      & EV-IMO pipeline \cite{8968520} & Event Frame & mIoU (\%) $\uparrow$ & 79.00 \\
    & & MSRNN \cite{zhang2023multi} & Voxel Grid & mIoU (\%) $\uparrow$ & 82.00 \\
    & & GConv \cite{mitrokhin2020learning} & Graph-based & mIoU (\%) $\uparrow$ & \textbf{87.00} \\
    & & PointNet++ \cite{mitrokhin2020learning} & Point-based & mIoU (\%) $\uparrow$ & 80.00 \\

 

 

\cmidrule{2-6}
 
    & \multirow{11}{*}{DSEC\cite{sun2022ess}} 
      & EV-SegNet \cite{alonso2019evsegnet} & Event Frame & mIoU (\%) $\uparrow$ & 51.76 \\
    & & HALSIE \cite{das2024halsie} & Event Frame & mIoU (\%) $\uparrow$ & 52.43 \\
    & & ESS (Event) \cite{sun2022ess} & Voxel Grid & mIoU (\%) $\uparrow$ & 51.57 \\
    & & ESS (Image-Event) \cite{sun2022ess} & Voxel Grid & mIoU (\%) $\uparrow$ & 53.29 \\
    & & EDCNet \cite{zhang2021exploring} & Voxel Grid & mIoU (\%) $\uparrow$ & 56.75 \\
    & & Hybrid-Seg \cite{li2025efficient} & Voxel Grid & mIoU (\%) $\uparrow$ & 66.57 \\
    & & MambaSeg \cite{gu2026mambaseg} & Voxel Grid & mIoU (\%) $\uparrow$ & \textbf{75.10} \\
    & & SE-Adapter \cite{yao2024sam} & Token-based & mIoU (\%) $\uparrow$ & 69.77 \\

\midrule
\multirow{24}{*}{\makecell[l]{Depth\\Estimation}} 
    & \multirow{8}{*}{EventScape\cite{gehrig2021ramnet}} 
      & EventDAM-L \cite{zhu2025depth} & Event Frame & 10/20/30m MAE $\downarrow$ & \textbf{0.56/1.52/2.30} \\
    & & EventDAM-B \cite{zhu2025depth} & Event Frame & 10/20/30m MAE $\downarrow$ & 0.54/1.54/2.32 \\
    & & EventDAM-S \cite{zhu2025depth} & Event Frame & 10/20/30m MAE $\downarrow$ & 0.59/1.60/2.47 \\
    & & SRFNet \cite{pan2024srfnet} &Voxel Grid  & 10/20/30m MAE $\downarrow$ & 1.27/1.68/2.76 \\
    & & RAMNet \cite{gehrig2021ramnet} & Voxel Grid & 10/20/30m MAE $\downarrow$ & 0.81/2.26/3.58 \\
    & & E2Depth \cite{hidalgo2020learning} & Voxel Grid & 10/20/30m MAE $\downarrow$ & 1.79/5.35/8.31 \\
    & & ER-F2D \cite{devulapally2024multi} & Token-based & 10/20/30m MAE $\downarrow$ & 0.67/1.69/2.81 \\

\cmidrule{2-6}
 
    & \multirow{5}{*}{DENSE\cite{hidalgo2020learning}} 
      & EventDAM-L \cite{zhu2025depth} & Event Frame & 10/20/30m MAE $\downarrow$ & \textbf{0.270/1.600/3.451} \\
    & & SRFNet \cite{pan2024srfnet} & Voxel Grid  & 10/20/30m MAE $\downarrow$ & 1.503/3.566/6.116 \\
    & & RAMNet \cite{gehrig2021ramnet} & Voxel Grid & 10/20/30m MAE $\downarrow$ & 2.619/11.264/19.113 \\

\cmidrule{2-6}
 
    & \multirow{9}{*}{MVSEC Night1\cite{zhu2018multivehicle}} 
      & Distil-E2D (E) \cite{lee2025distile2d} & Voxel Grid & 10/20/30m MAE $\downarrow$ & 1.41/2.16/2.75 \\
    & & Distil-E2D (E+I) \cite{lee2025distile2d} & Voxel Grid & 10/20/30m MAE $\downarrow$ & \textbf{1.35/2.02/2.79} \\
    & & RAMNet \cite{gehrig2021ramnet} & Voxel Grid & 10/20/30m MAE $\downarrow$ & 2.50/3.19/3.82 \\
    & & MDDE+ \cite{hidalgo2020learning} & Voxel Grid & 10/20/30m MAE $\downarrow$ & 3.38/3.82/4.46 \\
    & & ULODE \cite{zhu2019unsupervised} & Voxel Grid & 10/20/30m MAE $\downarrow$ & 3.13/4.02/4.89 \\
    & & EMoDepth \cite{zhu2023selfsupervised} & Voxel Grid & 10/20/30m MAE $\downarrow$ & 2.18/2.70/3.64 \\
    & & EvT+ (E+I) \cite{sabater2023event} & Token-based & 10/20/30m MAE $\downarrow$ & 1.45/2.10/2.88 \\
    & & EvT+ (E) \cite{sabater2023event} & Token-based & 10/20/30m MAE $\downarrow$ & 1.54/2.31/2.96 \\
    & & EReFormer \cite{liu2022eventdepth} & Token-based & 10/20/30m MAE $\downarrow$ & 1.85/2.45/3.24 \\

\cmidrule{2-6}
 
    & \multirow{2}{*}{DSEC\cite{gehrig2021dsec}} 
      & Distil-E2D (E) \cite{lee2025distile2d} & Voxel Grid & 10/20/30m MAE $\downarrow$ & \textbf{0.64/1.11/1.61} \\
    & & EMoDepth \cite{zhu2023selfsupervised} & Voxel Grid & 10/20/30m MAE $\downarrow$ & 0.96/1.55/2.21 \\

\midrule
\multirow{14}{*}{Optical Flow} 
    & \multirow{14}{*}{MVSEC\cite{zhu2018multivehicle}} 
      & STE-FlowNet \cite{ding2022spatio} & Event Frame & EPE/Out. $\downarrow$ & 0.42/0.00 \\
    & & EV-FlowNet \cite{zhu2018ev} & Time Surface & AEE/Out. $\downarrow$ & 0.49/0.20 \\
    & & DCR-EFlow \cite{sun2025dcr} & Voxel Grid & EPE/Out. $\downarrow$ & \textbf{0.20/0.00} \\
    & & DCEIFlow \cite{wan2022learning} & Voxel Grid & EPE/Out. $\downarrow$ & 0.22/0.00 \\
    & & E-RAFT \cite{gehrig2021eraft} & Voxel Grid & EPE/Out. $\downarrow$ & 0.24/0.00 \\
    & & TMA \cite{liu2023tma} & Voxel Grid & EPE/Out. $\downarrow$ & 0.25/0.07 \\
    & & EVA-Flow \cite{ye2025eva} & Voxel Grid & EPE/Out. $\downarrow$ & 0.25/0.00 \\
    & & Taming-CM \cite{shiba2023taming} & Voxel Grid & EPE/Out. $\downarrow$ & 0.27/0.05 \\
    & & Zhu et al. \cite{zhu2019unsupervised} & Voxel Grid & EPE/Out. $\downarrow$ & 0.32/0.00 \\
    & & ADM-Flow \cite{luo2023learning} & Voxel Grid & EPE/Out. $\downarrow$ & 0.41/0.00 \\
    & & Spike-FlowNet \cite{lee2020spikeflownet} & Voxel Grid & EPE $\downarrow$ & 0.49 \\
    & & Matrix-LSTM \cite{cannici2020differentiable} & Learned Dense & AEE/Out. $\downarrow$ & 0.821/0.471 \\
    & & ETB \cite{jiang2024token} & Token-based & AEE/Out. $\downarrow$ & 0.30/0.10 \\
    & & ET-FlowNet \cite{tian2022etflownet} & Token-based & EPE/Out. $\downarrow$ & 0.39/0.12 \\

\midrule
\multirow{11}{*}{Pose Estimation} 
    & \multirow{9}{*}{DHP19\cite{calabrese2019dhp19}} 
      & MAD-Det \cite{chen2025motion} & Dense--Dense & MPJPE (mm) $\downarrow$ & 56.32 \\
    & & TORE \cite{baldwin2022time} & Voxel Grid & MPJPE (mm) $\downarrow$ & 58.4 \\
    & & VMST-Net \cite{liu2023voxel} & Voxel Grid & MPJPE (mm) $\downarrow$ & 73.07 \\
    & & DGCNN \cite{chen2022efficient} & Graph-based & MPJPE (mm) $\downarrow$ & 77.32 \\
    & & HSPC \cite{tang2025event} & Point-based & MPJPE (mm) $\downarrow$ & \textbf{48.13} \\
    & & VMV-Point \cite{chen2022efficient} & Point-based & MPJPE (mm) $\downarrow$ & 103.23 \\
    & & SECNet \cite{ren2026scalable} & Point-based & MPJPE (mm) $\downarrow$ & 69.89 \\

\cmidrule{2-6}
 
    & \multirow{2}{*}{IJRR\cite{mueggler2017event}} 
      & EventMamba \cite{ren2024rethinking} & Point-based & CPR $\downarrow$ & \textbf{0.012} \\
    & & PEPNet \cite{ren2024simple} & Point-based & CPR $\downarrow$ & 0.013 \\

\midrule
\multirow{15}{*}{\makecell[l]{Low-level\\Vision}} 

 
    & \multirow{5}{*}{Event-Camera\cite{mueggler2017event}} 
      & TORE \cite{baldwin2022time} & Voxel Grid & SSIM $\uparrow$ & 0.550 \\
    & & CES \cite{lin2025compressed} & Voxel Grid & SSIM $\uparrow$ & 0.558 \\
    & & Sparse-E2VID \cite{cadena2023sparse} & Voxel Grid & SSIM $\uparrow$ & 0.606 \\
    & & E2VID \cite{rebecq2019high} & Voxel Grid & SSIM $\uparrow$ & \textbf{0.620} \\
    & & FireNet \cite{scheerlinck2020fast} & Voxel Grid & SSIM $\uparrow$ & \textbf{0.620} \\

 

 

\cmidrule{2-6}
 
    & \multirow{5}{*}{MVSEC\cite{zhu2018multivehicle}} 
      & E2VID \cite{rebecq2019high} & Voxel Grid & SSIM $\uparrow$ & 0.227 \\
    & & E2VID+ \cite{paredes2021back} & Voxel Grid & SSIM $\uparrow$ & 0.337 \\
    & & FireNet \cite{scheerlinck2020fast} & Voxel Grid & SSIM $\uparrow$ & 0.247 \\
    & & FireNet+ \cite{paredes2021back} & Voxel Grid & SSIM $\uparrow$ & 0.265 \\
    & & ET-Net \cite{weng2021event} & Token-based & SSIM $\uparrow$ & \textbf{0.358} \\

\cmidrule{2-6}
 
    & \multirow{5}{*}{GoPro\cite{nah2017deep}} 
      & EFNet \cite{han2022event} & Voxel Grid & PSNR/SSIM $\uparrow$ & 35.46/0.972 \\
    & & EIFNet \cite{yang2023event} & Voxel Grid & PSNR/SSIM $\uparrow$ & 35.99/0.979 \\
    & & MAENet \cite{sun2024motion} & Voxel Grid & PSNR/SSIM $\uparrow$ & 36.07/0.976 \\
    & & CFFNet \cite{li2024coarse} & Voxel Grid & PSNR/SSIM $\uparrow$ & 36.26/0.976 \\
    & & FEVD \cite{kim2024frequency} & Voxel Grid & PSNR/SSIM $\uparrow$ & \textbf{36.70/0.978} \\
    & & MAT \cite{xu2025motion} & Voxel Grid & PSNR/SSIM $\uparrow$ & 36.67/0.978 \\
    & & MTGNet \cite{lin2026event} & Dense--Sparse Hybrid & PSNR/SSIM $\uparrow$ &37.30/0.965\\

 

 

 

\end{longtable}
\twocolumn

\bibliographystyle{IEEEtran}
\bibliography{ieee}

@article{lagorce2016hots,
  title={Hots: a hierarchy of event-based time-surfaces for pattern recognition},
  author={Lagorce, Xavier and Orchard, Garrick and Galluppi, Francesco and Shi, Bertram E and Benosman, Ryad B},
  journal={IEEE transactions on pattern analysis and machine intelligence},
  volume={39},
  number={7},
  pages={1346--1359},
  year={2016},
  publisher={IEEE}
}

@inproceedings{liang2021event,
  title={Event-based object detection with lightweight spatial attention mechanism},
  author={Liang, Zichen and Chen, Guang and Li, Zhijun and Liu, Peigen and Knoll, Alois},
  booktitle={2021 6th IEEE International Conference on Advanced Robotics and Mechatronics (ICARM)},
  pages={498--503},
  year={2021},
  organization={IEEE}
}

@article{xu2025mets,
  title={METS: Motion-Encoded Time-Surface for Event-Based High-Speed Pose Tracking},
  author={Xu, Ninghui and Wang, Lihui and Yao, Zhiting and Okatani, Takayuki},
  journal={International Journal of Computer Vision},
  volume={133},
  number={7},
  pages={4401--4419},
  year={2025},
  publisher={Springer}
}

@inproceedings{zhu2023event,
  title={Event camera-based visual odometry for dynamic motion tracking of a legged robot using adaptive time surface},
  author={Zhu, Shifan and Tang, Zhipeng and Yang, Michael and Learned-Miller, Erik and Kim, Donghyun},
  booktitle={2023 IEEE/RSJ International Conference on Intelligent Robots and Systems (IROS)},
  pages={3475--3482},
  year={2023},
  organization={IEEE}
}

@inproceedings{sabater2022event,
  title={Event transformer. a sparse-aware solution for efficient event data processing},
  author={Sabater, Alberto and Montesano, Luis and Murillo, Ana C},
  booktitle={Proceedings of the IEEE/CVF Conference on Computer Vision and Pattern Recognition},
  pages={2677--2686},
  year={2022}
}

@inproceedings{zubic2023chaos,
  title={From chaos comes order: Ordering event representations for object recognition and detection},
  author={Zubi{\'c}, Nikola and Gehrig, Daniel and Gehrig, Mathias and Scaramuzza, Davide},
  booktitle={Proceedings of the IEEE/CVF International Conference on Computer Vision},
  pages={12846--12856},
  year={2023}
}

@inproceedings{manderscheid2019speed,
  title={Speed invariant time surface for learning to detect corner points with event-based cameras},
  author={Manderscheid, Jacques and Sironi, Amos and Bourdis, Nicolas and Migliore, Davide and Lepetit, Vincent},
  booktitle={Proceedings of the IEEE/CVF Conference on Computer Vision and Pattern Recognition},
  pages={10245--10254},
  year={2019}
}

@article{tang2026pa,
  title={PA-EVIO: Polarity-aided Event-Visual-Inertial Odometry with Adaptive Event Representation},
  author={Tang, Kai and Lang, Xiaolei and Ma, Yukai and Huang, Yuehao and Gu, Yaqing and Li, Laijian and Ren, Jie and Liu, Yong and Lv, Jiajun},
  journal={IEEE Transactions on Instrumentation and Measurement},
  year={2026},
  publisher={IEEE}
}

@inproceedings{chakravarthi2024recent,
  title={Recent event camera innovations: A survey},
  author={Chakravarthi, Bharatesh and Verma, Aayush Atul and Daniilidis, Kostas and Fermuller, Cornelia and Yang, Yezhou},
  booktitle={European conference on computer vision},
  pages={342--376},
  year={2024},
  organization={Springer}
}

@inproceedings{
ren2026scalable,
title={Scalable Event Cloud Network for Event-based Classification},
author={Ren, Hongwei and Ma, Fei and Lin, Xiaopeng and Fang, Yuetong and Huang, Hongxiang and Zhou, Yue and Huang, Yulong and Fu, Haotian and Yang, Ziyi and Jiang, Youxin and Wu, Xiangqian and Cheng, Bojun},
booktitle={Forty-third International Conference on Machine Learning},
year={2026},
url={https://openreview.net/forum?id=yAAUcDLYMR}
}

@inproceedings{ren2025e2b,
  title={E2B: A Single Modality Point-Based Tracker with Event Cameras},
  author={Ren, Hongwei and Li, Zhuo and Tuerhong, Aiersi and Liu, Haobo and Liang, Fei and Feng, Yongxiang and Wang, Wenhui and Wang, Yaoyuan and Zhang, Ziyang and He, Weihua and others},
  booktitle={2025 IEEE International Conference on Robotics and Automation (ICRA)},
  pages={6461--6468},
  year={2025},
  organization={IEEE}
}

@article{qi2017pointnet++,
  title={Pointnet++: Deep hierarchical feature learning on point sets in a metric space},
  author={Qi, Charles Ruizhongtai and Yi, Li and Su, Hao and Guibas, Leonidas J},
  journal={Advances in neural information processing systems},
  volume={30},
  year={2017}
}

@inproceedings{qi2017pointnet,
  title={Pointnet: Deep learning on point sets for 3d classification and segmentation},
  author={Qi, Charles R and Su, Hao and Mo, Kaichun and Guibas, Leonidas J},
  booktitle={Proceedings of the IEEE conference on computer vision and pattern recognition},
  pages={652--660},
  year={2017}
}

@article{perez2013mapping,
  title={Mapping from frame-driven to frame-free event-driven vision systems by low-rate rate coding and coincidence processing--application to feedforward ConvNets},
  author={P{\'e}rez-Carrasco, Jos{\'e} Antonio and Zhao, Bo and Serrano, Carmen and Acha, Begona and Serrano-Gotarredona, Teresa and Chen, Shouchun and Linares-Barranco, Bernab{\'e}},
  journal={IEEE transactions on pattern analysis and machine intelligence},
  volume={35},
  number={11},
  pages={2706--2719},
  year={2013},
  publisher={IEEE}
}

@article{zhao2014feedforward,
  title={Feedforward categorization on AER motion events using cortex-like features in a spiking neural network},
  author={Zhao, Bo and Ding, Ruoxi and Chen, Shoushun and Linares-Barranco, Bernabe and Tang, Huajin},
  journal={IEEE transactions on neural networks and learning systems},
  volume={26},
  number={9},
  pages={1963--1978},
  year={2014},
  publisher={IEEE}
}

@article{lee2016training,
  title={Training deep spiking neural networks using backpropagation},
  author={Lee, Jun Haeng and Delbruck, Tobi and Pfeiffer, Michael},
  journal={Frontiers in neuroscience},
  volume={10},
  pages={508},
  year={2016},
  publisher={Frontiers Media SA}
}

@inproceedings{rebecq2017real,
  title={Real-time Visual-Inertial Odometry for Event Cameras using Keyframe-based Nonlinear Optimization.},
  author={Rebecq, Henri and Horstschaefer, Timo and Scaramuzza, Davide},
  booktitle={BMVC},
  volume={2},
  number={3},
  pages={7},
  year={2017}
}

@article{zhu2018ev,
  title={EV-FlowNet: Self-supervised optical flow estimation for event-based cameras},
  author={Zhu, Alex Zihao and Yuan, Liangzhe and Chaney, Kenneth and Daniilidis, Kostas},
  journal={arXiv preprint arXiv:1802.06898},
  year={2018}
}

@inproceedings{huang2025exploring,
  title={Exploring temporal dynamics in event-based eye tracker},
  author={Huang, Hongxiang and Lin, Xiaopeng and Ren, Hongwei and Zhou, Yue and Cheng, Bojun},
  booktitle={Proceedings of the Computer Vision and Pattern Recognition Conference},
  pages={5145--5154},
  year={2025}
}

@inproceedings{wang2024mambapupil,
  title={Mambapupil: Bidirectional selective recurrent model for event-based eye tracking},
  author={Wang, Zhong and Wan, Zengyu and Han, Han and Liao, Bohao and Wu, Yuliang and Zhai, Wei and Cao, Yang and Zha, Zheng-Jun},
  booktitle={Proceedings of the IEEE/CVF Conference on Computer Vision and Pattern Recognition},
  pages={5762--5770},
  year={2024}
}

@article{wang2023visevent,
  title={Visevent: Reliable object tracking via collaboration of frame and event flows},
  author={Wang, Xiao and Li, Jianing and Zhu, Lin and Zhang, Zhipeng and Chen, Zhe and Li, Xin and Wang, Yaowei and Tian, Yonghong and Wu, Feng},
  journal={IEEE transactions on cybernetics},
  volume={54},
  number={3},
  pages={1997--2010},
  year={2023},
  publisher={IEEE}
}

@ARTICLE{4444573,
  author={Lichtsteiner, Patrick and Posch, Christoph and Delbruck, Tobi},
  journal={IEEE Journal of Solid-State Circuits}, 
  title={A 128$\times$ 128 120 dB 15 $\mu$s Latency Asynchronous Temporal Contrast Vision Sensor}, 
  year={2008},
  volume={43},
  number={2},
  pages={566-576},
  doi={10.1109/JSSC.2007.914337}}

@article{posch2010qvga,
  title={A QVGA 143 dB dynamic range frame-free PWM image sensor with lossless pixel-level video compression and time-domain CDS},
  author={Posch, Christoph and Matolin, Daniel and Wohlgenannt, Rainer},
  journal={IEEE Journal of Solid-State Circuits},
  volume={46},
  number={1},
  pages={259--275},
  year={2010},
  publisher={IEEE}
}

@article{chen2011efficient,
  title={Efficient feedforward categorization of objects and human postures with address-event image sensors},
  author={Chen, Shoushun and Akselrod, Polina and Zhao, Bo and Carrasco, Jose Antonio Perez and Linares-Barranco, Bernabe and Culurciello, Eugenio},
  journal={IEEE transactions on pattern analysis and machine intelligence},
  volume={34},
  number={2},
  pages={302--314},
  year={2011},
  publisher={IEEE}
}

@inproceedings{delbruck2007fast,
  title={Fast sensory motor control based on event-based hybrid neuromorphic-procedural system},
  author={Delbruck, Tobi and Lichtsteiner, Patrick},
  booktitle={2007 IEEE International Symposium on Circuits and Systems (ISCAS)},
  pages={845--848},
  year={2007},
  organization={IEEE}
}

@article{zheng2023deep,
  title={Deep learning for event-based vision: A comprehensive survey and benchmarks},
  author={Zheng, Xu and Liu, Yexin and Lu, Yunfan and Hua, Tongyan and Pan, Tianbo and Zhang, Weiming and Tao, Dacheng and Wang, Lin},
  journal={arXiv preprint arXiv:2302.08890},
  year={2023}
}

@article{gallego2020event,
  title={Event-based vision: A survey},
  author={Gallego, Guillermo and Delbr{\"u}ck, Tobi and Orchard, Garrick and Bartolozzi, Chiara and Taba, Brian and Censi, Andrea and Leutenegger, Stefan and Davison, Andrew J and Conradt, J{\"o}rg and Daniilidis, Kostas and others},
  journal={IEEE transactions on pattern analysis and machine intelligence},
  volume={44},
  number={1},
  pages={154--180},
  year={2020},
  publisher={IEEE}
}

@article{shariff2024event,
  title={Event cameras in automotive sensing: A review},
  author={Shariff, Waseem and Dilmaghani, Mehdi Sefidgar and Kielty, Paul and Moustafa, Mohamed and Lemley, Joe and Corcoran, Peter},
  journal={IEEE Access},
  volume={12},
  pages={51275--51306},
  year={2024},
  publisher={IEEE}
}

@article{benosman2013event,
  title={Event-based visual flow},
  author={Benosman, Ryad and Clercq, Charles and Lagorce, Xavier and Ieng, Sio-Hoi and Bartolozzi, Chiara},
  journal={IEEE transactions on neural networks and learning systems},
  volume={25},
  number={2},
  pages={407--417},
  year={2013},
  publisher={IEEE}
}

@inproceedings{sironi2018hats,
  title={HATS: Histograms of averaged time surfaces for robust event-based object classification},
  author={Sironi, Amos and Brambilla, Manuele and Bourdis, Nicolas and Lagorce, Xavier and Benosman, Ryad},
  booktitle={Proceedings of the IEEE conference on computer vision and pattern recognition},
  pages={1731--1740},
  year={2018}
}

@article{almatrafi2020distance,
  title={Distance surface for event-based optical flow},
  author={Almatrafi, Mohammed and Baldwin, Raymond and Aizawa, Kiyoharu and Hirakawa, Keigo},
  journal={IEEE transactions on pattern analysis and machine intelligence},
  volume={42},
  number={7},
  pages={1547--1556},
  year={2020},
  publisher={IEEE}
}

@article{chen2022ecsnet,
  title={ECSNet: Spatio-temporal feature learning for event camera},
  author={Chen, Zhiwen and Wu, Jinjian and Hou, Junhui and Li, Leida and Dong, Weisheng and Shi, Guangming},
  journal={IEEE Transactions on Circuits and Systems for Video Technology},
  volume={33},
  number={2},
  pages={701--712},
  year={2022},
  publisher={IEEE}
}

@inproceedings{liu2024video,
  title={Video frame interpolation via direct synthesis with the event-based reference},
  author={Liu, Yuhan and Deng, Yongjian and Chen, Hao and Yang, Zhen},
  booktitle={Proceedings of the IEEE/CVF Conference on Computer Vision and Pattern Recognition},
  pages={8477--8487},
  year={2024}
}

@article{mueggler2017fast,
  title={Fast event-based corner detection},
  author={Mueggler, Elias and Bartolozzi, Chiara and Scaramuzza, Davide},
  year={2017},
  publisher={University of Zurich}
}

@inproceedings{zhu2019unsupervised,
  title={Unsupervised event-based learning of optical flow, depth, and egomotion},
  author={Zhu, Alex Zihao and Yuan, Liangzhe and Chaney, Kenneth and Daniilidis, Kostas},
  booktitle={Proceedings of the IEEE/CVF conference on computer vision and pattern recognition},
  pages={989--997},
  year={2019}
}

@article{rebecq2019high,
  title={High speed and high dynamic range video with an event camera},
  author={Rebecq, Henri and Ranftl, Ren{\'e} and Koltun, Vladlen and Scaramuzza, Davide},
  journal={IEEE transactions on pattern analysis and machine intelligence},
  volume={43},
  number={6},
  pages={1964--1980},
  year={2019},
  publisher={IEEE}
}

@article{perot2020learning,
  title={Learning to detect objects with a 1 megapixel event camera},
  author={Perot, Etienne and De Tournemire, Pierre and Nitti, Davide and Masci, Jonathan and Sironi, Amos},
  journal={Advances in Neural Information Processing Systems},
  volume={33},
  pages={16639--16652},
  year={2020}
}

@article{baldwin2022time,
  title={Time-ordered recent event (tore) volumes for event cameras},
  author={Baldwin, R Wes and Liu, Ruixu and Almatrafi, Mohammed and Asari, Vijayan and Hirakawa, Keigo},
  journal={IEEE Transactions on Pattern Analysis and Machine Intelligence},
  volume={45},
  number={2},
  pages={2519--2532},
  year={2022},
  publisher={IEEE}
}

@inproceedings{zhao2022transformer,
  title={Transformer-based domain adaptation for event data classification},
  author={Zhao, Junwei and Zhang, Shiliang and Huang, Tiejun},
  booktitle={ICASSP 2022-2022 IEEE International Conference on Acoustics, Speech and Signal Processing (ICASSP)},
  pages={4673--4677},
  year={2022},
  organization={IEEE}
}

@article{liu2022fast,
  title={Fast classification and action recognition with event-based imaging},
  author={Liu, Chang and Qi, Xiaojuan and Lam, Edmund Y and Wong, Ngai},
  journal={IEEE access},
  volume={10},
  pages={55638--55649},
  year={2022},
  publisher={IEEE}
}

@inproceedings{cadena2023sparse,
  title={Sparse-e2vid: A sparse convolutional model for event-based video reconstruction trained with real event noise},
  author={Cadena, Pablo Rodrigo Gantier and Qian, Yeqiang and Wang, Chunxiang and Yang, Ming},
  booktitle={Proceedings of the IEEE/CVF Conference on Computer Vision and Pattern Recognition},
  pages={4150--4158},
  year={2023}
}

@article{zhang2023multi,
  title={A multi-scale recurrent framework for motion segmentation with event camera},
  author={Zhang, Shaobo and Sun, Lei and Wang, Kaiwei},
  journal={IEEE Access},
  volume={11},
  pages={80105--80114},
  year={2023},
  publisher={IEEE}
}

@inproceedings{luo2023learning,
  title={Learning optical flow from event camera with rendered dataset},
  author={Luo, Xinglong and Luo, Kunming and Luo, Ao and Wang, Zhengning and Tan, Ping and Liu, Shuaicheng},
  booktitle={Proceedings of the IEEE/CVF International Conference on Computer Vision},
  pages={9847--9857},
  year={2023}
}

@article{lin2025compressed,
  title={Compressed event sensing (CES) volumes for event cameras},
  author={Lin, Songnan and Ma, Ye and Chen, Jing and Wen, Bihan},
  journal={International Journal of Computer Vision},
  volume={133},
  number={1},
  pages={435--455},
  year={2025},
  publisher={Springer}
}

@article{ren2023spikepoint,
  title={Spikepoint: An efficient point-based spiking neural network for event cameras action recognition},
  author={Ren, Hongwei and Zhou, Yue and Huang, Yulong and Fu, Haotian and Lin, Xiaopeng and Song, Jie and Cheng, Bojun},
  journal={arXiv preprint arXiv:2310.07189},
  year={2023}
}

@inproceedings{yao2024sam,
  title={Sam-event-adapter: Adapting segment anything model for event-rgb semantic segmentation},
  author={Yao, Bowen and Deng, Yongjian and Liu, Yuhan and Chen, Hao and Li, Youfu and Yang, Zhen},
  booktitle={2024 IEEE International Conference on Robotics and Automation (ICRA)},
  pages={9093--9100},
  year={2024},
  organization={IEEE}
}

@article{xie2024event,
  title={Event voxel set transformer for spatiotemporal representation learning on event streams},
  author={Xie, Bochen and Deng, Yongjian and Shao, Zhanpeng and Xu, Qingsong and Li, Youfu},
  journal={IEEE Transactions on Circuits and Systems for Video Technology},
  volume={34},
  number={12},
  pages={13427--13440},
  year={2024},
  publisher={IEEE}
}

@article{liu2023voxel,
  title={Voxel-based multi-scale transformer network for event stream processing},
  author={Liu, Daikun and Wang, Teng and Sun, Changyin},
  journal={IEEE Transactions on Circuits and Systems for Video Technology},
  volume={34},
  number={4},
  pages={2112--2124},
  year={2023},
  publisher={IEEE}
}

@article{zhu2024continuous,
  title={Continuous-time object segmentation using high temporal resolution event camera},
  author={Zhu, Lin and Chen, Xianzhang and Wang, Lizhi and Wang, Xiao and Tian, Yonghong and Huang, Hua},
  journal={IEEE Transactions on Pattern Analysis and Machine Intelligence},
  year={2024},
  publisher={IEEE}
}

@inproceedings{wang2024event,
  title={Event stream-based visual object tracking: A high-resolution benchmark dataset and a novel baseline},
  author={Wang, Xiao and Wang, Shiao and Tang, Chuanming and Zhu, Lin and Jiang, Bo and Tian, Yonghong and Tang, Jin},
  booktitle={Proceedings of the IEEE/CVF Conference on Computer Vision and Pattern Recognition},
  pages={19248--19257},
  year={2024}
}

@inproceedings{barchid2022bina,
  title={Bina-rep event frames: A simple and effective representation for event-based cameras},
  author={Barchid, Sami and Mennesson, Jos{\'e} and Dj{\'e}raba, Chaabane},
  booktitle={2022 IEEE International Conference on Image Processing (ICIP)},
  pages={3998--4002},
  year={2022},
  organization={IEEE}
}

@INPROCEEDINGS{9412991,
  author={Innocenti, Simone Undri and Becattini, Federico and Pernici, Federico and Del Bimbo, Alberto},
  booktitle={2020 25th International Conference on Pattern Recognition (ICPR)}, 
  title={Temporal Binary Representation for Event-Based Action Recognition}, 
  year={2021},
  volume={},
  number={},
  pages={10426-10432},
  doi={10.1109/ICPR48806.2021.9412991}}

@inproceedings{wu2025eventmg,
title={Event{MG}: Efficient Multilevel Mamba-Graph Learning for Spatiotemporal Event Representation},
author={Sheng Wu and Lin Jin and Hui Feng and Bo Hu},
booktitle={The Thirty-ninth Annual Conference on Neural Information Processing Systems},
year={2025},
url={https://openreview.net/forum?id=OmtKcee8NA}
}

@article{bi2020graph,
  title={Graph-based spatio-temporal feature learning for neuromorphic vision sensing},
  author={Bi, Yin and Chadha, Aaron and Abbas, Alhabib and Bourtsoulatze, Eirina and Andreopoulos, Yiannis},
  journal={IEEE Transactions on Image Processing},
  volume={29},
  pages={9084--9098},
  year={2020},
  publisher={IEEE}
}

@inproceedings{li2021graph,
  title={Graph-based asynchronous event processing for rapid object recognition},
  author={Li, Yijin and Zhou, Han and Yang, Bangbang and Zhang, Ye and Cui, Zhaopeng and Bao, Hujun and Zhang, Guofeng},
  booktitle={Proceedings of the IEEE/CVF International Conference on Computer Vision},
  pages={934--943},
  year={2021}
}

@article{zhu2022learning,
  title={Learning graph-embedded key-event back-tracing for object tracking in event clouds},
  author={Zhu, Zhiyu and Hou, Junhui and Lyu, Xianqiang},
  journal={Advances in Neural Information Processing Systems},
  volume={35},
  pages={7462--7476},
  year={2022}
}

@article{xie2022vmv,
  title={Vmv-gcn: Volumetric multi-view based graph cnn for event stream classification},
  author={Xie, Bochen and Deng, Yongjian and Shao, Zhanpeng and Liu, Hai and Li, Youfu},
  journal={IEEE Robotics and Automation Letters},
  volume={7},
  number={2},
  pages={1976--1983},
  year={2022},
  publisher={IEEE}
}

@inproceedings{deng2022voxel,
  title={A voxel graph CNN for object classification with event cameras},
  author={Deng, Yongjian and Chen, Hao and Liu, Hai and Li, Youfu},
  booktitle={Proceedings of the IEEE/CVF Conference on Computer Vision and Pattern Recognition},
  pages={1172--1181},
  year={2022}
}

@inproceedings{schaefer2022aegnn,
  title={Aegnn: Asynchronous event-based graph neural networks},
  author={Schaefer, Simon and Gehrig, Daniel and Scaramuzza, Davide},
  booktitle={Proceedings of the IEEE/CVF conference on computer vision and pattern recognition},
  pages={12371--12381},
  year={2022}
}

@inproceedings{yuan2023learning,
  title={Learning bottleneck transformer for event image-voxel feature fusion based classification},
  author={Yuan, Chengguo and Jin, Yu and Wu, Zongzhen and Wei, Fanting and Wang, Yangzirui and Chen, Lan and Wang, Xiao},
  booktitle={Chinese Conference on Pattern Recognition and Computer Vision (PRCV)},
  pages={3--15},
  year={2023},
  organization={Springer}
}

@inproceedings{deng2024dynamic,
  title={A dynamic GCN with cross-representation distillation for event-based learning},
  author={Deng, Yongjian and Chen, Hao and Li, Youfu},
  booktitle={Proceedings of the AAAI Conference on Artificial Intelligence},
  volume={38},
  number={2},
  pages={1492--1500},
  year={2024}
}

@inproceedings{fan2025eventpillars,
  title={Eventpillars: Pillar-based efficient representations for event data},
  author={Fan, Rui and Hao, Weidong and Guan, Juntao and Rui, Lai and Gu, Lin and Wu, Tong and Zeng, Fanhong and Zhu, Zhangming},
  booktitle={Proceedings of the AAAI Conference on Artificial Intelligence},
  volume={39},
  number={3},
  pages={2861--2869},
  year={2025}
}

@inproceedings{gehrig2019end,
  title={End-to-end learning of representations for asynchronous event-based data},
  author={Gehrig, Daniel and Loquercio, Antonio and Derpanis, Konstantinos G and Scaramuzza, Davide},
  booktitle={Proceedings of the IEEE/CVF international conference on computer vision},
  pages={5633--5643},
  year={2019}
}

@inproceedings{cannici2020differentiable,
  title={A differentiable recurrent surface for asynchronous event-based data},
  author={Cannici, Marco and Ciccone, Marco and Romanoni, Andrea and Matteucci, Matteo},
  booktitle={European Conference on Computer Vision},
  pages={136--152},
  year={2020},
  organization={Springer}
}

@inproceedings{peng2023get,
  title={Get: Group event transformer for event-based vision},
  author={Peng, Yansong and Zhang, Yueyi and Xiong, Zhiwei and Sun, Xiaoyan and Wu, Feng},
  booktitle={Proceedings of the IEEE/CVF International Conference on Computer Vision},
  pages={6038--6048},
  year={2023}
}

@inproceedings{jiang2024token,
  title={Token-based spatiotemporal representation of the events},
  author={Jiang, Bin and Li, Zhihao and Asif, M Salman and Cao, Xun and Ma, Zhan},
  booktitle={ICASSP 2024-2024 IEEE International Conference on Acoustics, Speech and Signal Processing (ICASSP)},
  pages={5240--5244},
  year={2024},
  organization={IEEE}
}

@inproceedings{sekikawa2019eventnet,
  title={Eventnet: Asynchronous recursive event processing},
  author={Sekikawa, Yusuke and Hara, Kosuke and Saito, Hideo},
  booktitle={Proceedings of the IEEE/CVF conference on computer vision and pattern recognition},
  pages={3887--3896},
  year={2019}
}

@inproceedings{wang2019space,
  title={Space-time event clouds for gesture recognition: From RGB cameras to event cameras},
  author={Wang, Qinyi and Zhang, Yexin and Yuan, Junsong and Lu, Yilong},
  booktitle={2019 IEEE Winter Conference on Applications of Computer Vision (WACV)},
  pages={1826--1835},
  year={2019},
  organization={IEEE}
}

@inproceedings{yang2019modeling,
  title={Modeling point clouds with self-attention and gumbel subset sampling},
  author={Yang, Jiancheng and Zhang, Qiang and Ni, Bingbing and Li, Linguo and Liu, Jinxian and Zhou, Mengdie and Tian, Qi},
  booktitle={Proceedings of the IEEE/CVF conference on computer vision and pattern recognition},
  pages={3323--3332},
  year={2019}
}

@inproceedings{chen2022efficient,
  title={Efficient human pose estimation via 3d event point cloud},
  author={Chen, Jiaan and Shi, Hao and Ye, Yaozu and Yang, Kailun and Sun, Lei and Wang, Kaiwei},
  booktitle={2022 International Conference on 3D Vision (3DV)},
  pages={1--10},
  year={2022},
  organization={IEEE}
}

@inproceedings{ren2023ttpoint,
  title={Ttpoint: A tensorized point cloud network for lightweight action recognition with event cameras},
  author={Ren, Hongwei and Zhou, Yue and Fu, Haotian and Huang, Yulong and Xu, Renjing and Cheng, Bojun},
  booktitle={Proceedings of the 31st ACM International Conference on Multimedia},
  pages={8026--8034},
  year={2023}
}

@inproceedings{ren2024simple,
  title={A simple and effective point-based network for event camera 6-dofs pose relocalization},
  author={Ren, Hongwei and Zhu, Jiadong and Zhou, Yue and Fu, Haotian and Huang, Yulong and Cheng, Bojun},
  booktitle={Proceedings of the IEEE/CVF Conference on Computer Vision and Pattern Recognition},
  pages={18112--18121},
  year={2024}
}

@article{ren2024rethinking,
  title={Rethinking efficient and effective point-based networks for event camera classification and regression: Eventmamba},
  author={Ren, Hongwei and Zhou, Yue and Zhu, Jiadong and Fu, Haotian and Huang, Yulong and Lin, Xiaopeng and Fang, Yuetong and Ma, Fei and Yu, Hao and Cheng, Bojun},
  journal={arXiv preprint arXiv:2405.06116},
  year={2024}
}

@inproceedings{zheng2024eventdance,
  title={Eventdance: Unsupervised source-free cross-modal adaptation for event-based object recognition},
  author={Zheng, Xu and Wang, Lin},
  booktitle={Proceedings of the IEEE/CVF Conference on Computer Vision and Pattern Recognition},
  pages={17448--17458},
  year={2024}
}

@article{chen2025motion,
  title={Motion and appearance decoupling representation for event cameras},
  author={Chen, Nuo and Li, Boyang and Wang, Yingqian and Ying, Xinyi and Wang, Longguang and Zhang, Chushu and Guo, Yulan and Li, Miao and An, Wei},
  journal={IEEE Transactions on Image Processing},
  year={2025},
  publisher={IEEE}
}

@inproceedings{amir2017low,
  title={A low power, fully event-based gesture recognition system},
  author={Amir, Arnon and Taba, Brian and Berg, David and Melano, Timothy and McKinstry, Jeffrey and Di Nolfo, Carmelo and Nayak, Tapan and Andreopoulos, Alexander and Garreau, Guillaume and Mendoza, Marcela and others},
  booktitle={Proceedings of the IEEE conference on computer vision and pattern recognition},
  pages={7243--7252},
  year={2017}
}

@article{wang2019dynamic,
  title={Dynamic graph cnn for learning on point clouds},
  author={Wang, Yue and Sun, Yongbin and Liu, Ziwei and Sarma, Sanjay E and Bronstein, Michael M and Solomon, Justin M},
  journal={ACM Transactions on Graphics (tog)},
  volume={38},
  number={5},
  pages={1--12},
  year={2019},
  publisher={Acm New York, NY, USA}
}

@article{orchard2015converting,
  title={Converting static image datasets to spiking neuromorphic datasets using saccades},
  author={Orchard, Garrick and Jayawant, Ajinkya and Cohen, Gregory K and Thakor, Nitish},
  journal={Frontiers in neuroscience},
  volume={9},
  pages={437},
  year={2015},
  publisher={Frontiers Media SA}
}

@article{wang2024dailydvs,
  title={DailyDVS-200: A Comprehensive Benchmark Dataset for Event-Based Action Recognition},
  author={Wang, Qi and Xu, Zhou and Lin, Yuming and Ye, Jingtao and Li, Hongsheng and Zhu, Guangming and Shah, Syed Afaq Ali and Bennamoun, Mohammed and Zhang, Liang},
  journal={arXiv preprint arXiv:2407.05106},
  year={2024}
}

@article{binas2017ddd17,
  title={DDD17: End-to-end DAVIS driving dataset},
  author={Binas, Jonathan and Neil, Daniel and Liu, Shih-Chii and Delbruck, Tobi},
  journal={arXiv preprint arXiv:1711.01458},
  year={2017}
}

@article{gao2023action,
  title={Action recognition and benchmark using event cameras},
  author={Gao, Yue and Lu, Jiaxuan and Li, Siqi and Ma, Nan and Du, Shaoyi and Li, Yipeng and Dai, Qionghai},
  journal={IEEE Transactions on Pattern Analysis and Machine Intelligence},
  volume={45},
  number={12},
  pages={14081--14097},
  year={2023},
  publisher={IEEE}
}

@article{cao2024eventcrab,
  title={Eventcrab: Harnessing frame and point synergy for event-based action recognition and beyond},
  author={Cao, Meiqi and Shu, Xiangbo and Zhang, Jiachao and Yan, Rui and Li, Zechao and Tang, Jinhui},
  journal={arXiv preprint arXiv:2411.18328},
  year={2024}
}

@article{lin2026event,
  title={Event-based Motion Deblurring via Multi-Temporal Granularity Fusion},
  author={Lin, Xiaopeng and Ren, Hongwei and Huang, Yulong and Liu, Zunchang and Zhou, Yue and Fu, Haotian and Pan, Biao and Cheng, Bojun},
  journal={IEEE Transactions on Circuits and Systems for Video Technology},
  year={2026},
  publisher={IEEE}
}

@inproceedings{zhang2025mtga,
  title={Mtga: Multi-view temporal granularity aligned aggregation for event-based lip-reading},
  author={Zhang, Wenhao and Wang, Jun and Luo, Yong and Yu, Lei and Yu, Wei and He, Zheng and Shen, Jialie},
  booktitle={Proceedings of the AAAI Conference on Artificial Intelligence},
  volume={39},
  number={10},
  pages={10176--10184},
  year={2025}
}

@inproceedings{fan2026smv,
  title={SMV-EAR: Bring Spatiotemporal Multi-View Representation Learning into Efficient Event-Based Action Recognition},
  author={Fan, Rui and Hao, Weidong and Guan, Juntao and Rui, Lai and Wu, Tong and Zeng, Fanhong and Gu, Lin},
  booktitle={Proceedings of the IEEE/CVF Conference on Computer Vision and Pattern Recognition},
  pages={6043--6053},
  year={2026}
}

@article{chen2025retain,
  title={Retain, blend, and exchange: A quality-aware spatial-stereo fusion approach for event stream recognition},
  author={Chen, Lan and Li, Dong and Wang, Xiao and Shao, Pengpeng and Zhang, Wei and Wang, Yaowei and Tian, Yonghong and Tang, Jin},
  journal={IEEE Transactions on Multimedia},
  volume={27},
  pages={8926--8939},
  year={2025},
  publisher={IEEE}
}

@article{liu2026spike,
  title={Spike-EVPR: Deep Spiking Residual Networks with SNN-Tailored Representations for Event-Based Visual Place Recognition},
  author={Liu, Zuntao and Li, Yaohui and Hu, Chenming and Kong, Delei and Jiang, Junjie and Fang, Zheng},
  journal={IEEE Robotics and Automation Letters},
  year={2026},
  publisher={IEEE}
}

@article{de2020large,
  title={A large scale event-based detection dataset for automotive},
  author={De Tournemire, Pierre and Nitti, Davide and Perot, Etienne and Migliore, Davide and Sironi, Amos},
  journal={arXiv preprint arXiv:2001.08499},
  year={2020}
}

@INPROCEEDINGS{8968520,
  author={Mitrokhin, Anton and Ye, Chengxi and Fermüller, Cornelia and Aloimonos, Yiannis and Delbruck, Tobi},
  booktitle={2019 IEEE/RSJ International Conference on Intelligent Robots and Systems (IROS)}, 
  title={EV-IMO: Motion Segmentation Dataset and Learning Pipeline for Event Cameras}, 
  year={2019},
  volume={},
  number={},
  pages={6105-6112},
  doi={10.1109/IROS40897.2019.8968520}}

@article{zhu2018multivehicle,
  title={The multivehicle stereo event camera dataset: An event camera dataset for 3D perception},
  author={Zhu, Alex Zihao and Thakur, Dinesh and {\"O}zaslan, Tolga and Pfrommer, Bernd and Kumar, Vijay and Daniilidis, Kostas},
  journal={IEEE Robotics and Automation Letters},
  volume={3},
  number={3},
  pages={2032--2039},
  year={2018},
  publisher={IEEE}
}

@inproceedings{calabrese2019dhp19,
  title={DHP19: Dynamic vision sensor 3D human pose dataset},
  author={Calabrese, Enrico and Taverni, Gemma and Awai Easthope, Christopher and Skriabine, Sophie and Corradi, Federico and Longinotti, Luca and Eng, Kynan and Delbruck, Tobi},
  booktitle={Proceedings of the IEEE/CVF conference on computer vision and pattern recognition workshops},
  pages={0--0},
  year={2019}
}

@article{mueggler2017event,
  title={The event-camera dataset and simulator: Event-based data for pose estimation, visual odometry, and SLAM},
  author={Mueggler, Elias and Rebecq, Henri and Gallego, Guillermo and Delbr{\"u}ck, Tobi and Scaramuzza, Davide},
  journal={The International Journal of Robotics Research},
  volume={36},
  number={2},
  pages={145--149},
  year={2017},
  publisher={SAGE Publications Sage UK: London, England}
}

@article{wang2025event,
  title={Event stream-based visual object tracking: HDETrack V2 and a high-definition benchmark},
  author={Wang, Shiao and Wang, Xiao and Wang, Chao and Jin, Liye and Zhu, Lin and Jiang, Bo and Tian, Yonghong and Tang, Jin},
  journal={arXiv preprint arXiv:2502.05574},
  year={2025}
}

@article{zhu2025crsot,
  title={Crsot: Cross-resolution object tracking using unaligned frame and event cameras},
  author={Zhu, Yabin and Wang, Xiao and Li, Chenglong and Jiang, Bo and Zhu, Lin and Huang, Zhixiang and Tian, Yonghong and Tang, Jin},
  journal={IEEE Transactions on Multimedia},
  year={2025},
  publisher={IEEE}
}

@article{wang2025mambaevt,
  title={Mambaevt: Event stream based visual object tracking using state space model},
  author={Wang, Xiao and Wang, Chao and Wang, Shiao and Wang, Xixi and Zhao, Zhicheng and Zhu, Lin and Jiang, Bo},
  journal={IEEE Transactions on Circuits and Systems for Video Technology},
  year={2025},
  publisher={IEEE}
}

@article{wang2025mamba,
  title={Mamba-FETrack V2: Revisiting State Space Model for Frame-Event based Visual Object Tracking},
  author={Wang, Shiao and Huang, Ju and Ma, Qingchuan and Gao, Jinfeng and Xu, Chunyi and Wang, Xiao and Chen, Lan and Jiang, Bo},
  journal={arXiv preprint arXiv:2506.23783},
  year={2025}
}

@article{wang2025towards,
  title={Towards low-latency event stream-based visual object tracking: A slow-fast approach},
  author={Wang, Shiao and Wang, Xiao and Jin, Liye and Jiang, Bo and Zhu, Lin and Chen, Lan and Tian, Yonghong and Luo, Bin},
  journal={arXiv preprint arXiv:2505.12903},
  year={2025}
}

@article{tang2025revisiting,
  title={Revisiting color-event based tracking: A unified network, dataset, and metric},
  author={Tang, Chuanming and Wang, Xiao and Huang, Ju and Jiang, Bo and Zhu, Lin and Chen, Shifeng and Zhang, Jianlin and Wang, Yaowei and Tian, Yonghong},
  journal={Pattern Recognition},
  pages={112718},
  year={2025},
  publisher={Elsevier}
}

@inproceedings{zhang2022spiking,
  title={Spiking transformers for event-based single object tracking},
  author={Zhang, Jiqing and Dong, Bo and Zhang, Haiwei and Ding, Jianchuan and Heide, Felix and Yin, Baocai and Yang, Xin},
  booktitle={Proceedings of the IEEE/CVF conference on Computer Vision and Pattern Recognition},
  pages={8801--8810},
  year={2022}
}

@inproceedings{huang2024mamba,
  title={Mamba-fetrack: Frame-event tracking via state space model},
  author={Huang, Ju and Wang, Shiao and Wang, Shuai and Wu, Zhe and Wang, Xiao and Jiang, Bo},
  booktitle={Chinese Conference on Pattern Recognition and Computer Vision (PRCV)},
  pages={3--18},
  year={2024},
  organization={Springer}
}

@inproceedings{zhang2021object,
  title={Object tracking by jointly exploiting frame and event domain},
  author={Zhang, Jiqing and Yang, Xin and Fu, Yingkai and Wei, Xiaopeng and Yin, Baocai and Dong, Bo},
  booktitle={Proceedings of the IEEE/CVF international conference on computer vision},
  pages={13043--13052},
  year={2021}
}

@inproceedings{he2022timereplayer,
  title={TimeReplayer: Unlocking the Potential of Event Cameras for Video Interpolation},
  author={He, Weihua and You, Kaichao and Qiao, Zhendong and Jia, Xu and Zhang, Ziyang and Wang, Wenhui and Lu, Huchuan and Wang, Yaoyuan and Liao, Jianxing},
  booktitle={Proceedings of the IEEE/CVF Conference on Computer Vision and Pattern Recognition},
  pages={17783--17792},
  year={2022}
}

@inproceedings{tulyakov2021timelens,
  title={Time Lens: Event-based Video Frame Interpolation},
  author={Tulyakov, Stepan and Gehrig, Daniel and Georgoulis, Stamatios and Erbach, Julius and Gehrig, Mathias and Li, Yuanyou and Scaramuzza, Davide},
  booktitle={Proceedings of the IEEE/CVF Conference on Computer Vision and Pattern Recognition},
  pages={16155--16164},
  year={2021}
}

@inproceedings{kim2023cbmnet,
  title={Event-Based Video Frame Interpolation with Cross-Modal Asymmetric Bidirectional Motion Fields},
  author={Kim, Taewoo and Chae, Yujeong and Jang, Hyun-Kurl and Yoon, Kuk-Jin},
  booktitle={Proceedings of the IEEE/CVF Conference on Computer Vision and Pattern Recognition},
  pages={18032--18042},
  year={2023}
}

@inproceedings{nah2017deep,
  title={Deep Multi-Scale Convolutional Neural Network for Dynamic Scene Deblurring},
  author={Nah, Seungjun and Kim, Tae Hyun and Lee, Kyoung Mu},
  booktitle={Proceedings of the IEEE Conference on Computer Vision and Pattern Recognition},
  pages={3883--3891},
  year={2017}
}

@inproceedings{liu2021eventaction,
  title={Event-based Action Recognition Using Motion Information and Spiking Neural Networks},
  author={Liu, Qianhui and Xing, Dong and Tang, Huajin and Ma, De and Pan, Gang},
  booktitle={Proceedings of the Thirtieth International Joint Conference on Artificial Intelligence},
  pages={1743--1749},
  year={2021},
  doi={10.24963/ijcai.2021/240}
}

@inproceedings{shrestha2018slayer,
  title={SLAYER: Spike Layer Error Reassignment in Time},
  author={Shrestha, Sumit Bam and Orchard, Garrick},
  booktitle={Advances in Neural Information Processing Systems},
  volume={31},
  year={2018}
}

@inproceedings{yao2021temporal,
  title={Temporal-Wise Attention Spiking Neural Networks for Event Streams Classification},
  author={Yao, Man and Gao, Huanhuan and Zhao, Guangshe and Wang, Dingheng and Lin, Yihan and Yang, Zhaoxu and Li, Guoqi},
  booktitle={Proceedings of the IEEE/CVF International Conference on Computer Vision},
  pages={10221--10230},
  year={2021}
}

@article{xu2023stsc,
  title={STSC-SNN: Spatio-Temporal Synaptic Connection with Temporal Convolution and Attention for Spiking Neural Networks},
  author={Xu, Qi and Li, Yaxin and Shen, Jiangrong and Liu, Jian K. and Tang, Huajin and Pan, Gang},
  journal={Frontiers in Neuroscience},
  volume={16},
  pages={1079357},
  year={2023},
  doi={10.3389/fnins.2022.1079357}
}

@article{zhu2022tcja,
  author={Zhu, Rui-Jie and Zhang, Malu and Zhao, Qihang and Deng, Haoyu and Duan, Yule and Deng, Liang-Jian},
  title={TCJA-SNN: Temporal-Channel Joint Attention for Spiking Neural Networks},
  journal={arXiv preprint arXiv:2206.10177},
  year={2022}
}

@article{xiao2019hmax,
  title={An Event-Driven Categorization Model for AER Image Sensors Using Multispike Encoding and Learning},
  author={Xiao, Rong and Tang, Huajin and Ma, Yuhao and Yan, Rui and Orchard, Garrick},
  journal={IEEE Transactions on Neural Networks and Learning Systems},
  year={2019}
}

@inproceedings{liu2020event,
  title={Effective AER Object Classification Using Segmented Probability-Maximization Learning in Spiking Neural Networks},
  author={Liu, Qianhui and Ruan, Haibo and Xing, Dong and Tang, Huajin and Pan, Gang},
  booktitle={Proceedings of the AAAI Conference on Artificial Intelligence},
  pages={1308--1315},
  year={2020}
}

@inproceedings{gao2024hardvs,
  title={HARDVS: Revisiting Human Activity Recognition with Dynamic Vision Sensors},
  author={Gao, Yue and Lu, Jiapeng and Li, Siqi and Ma, Nan and Du, Sheng and Li, Yipeng and Dai, Qionghai},
  booktitle={Proceedings of the AAAI Conference on Artificial Intelligence},
  year={2024}
}

@inproceedings{cao2024embracing,
  title={Embracing Events and Frames with Hierarchical Feature Refinement Network for Object Detection},
  author={Cao, Hu and Zhang, Zehua and Xia, Yan and Li, Xinyi and Xia, Jiahao and Chen, Guang and Knoll, Alois},
  booktitle={European Conference on Computer Vision},
  pages={1--17},
  year={2024},
  organization={Springer}
}

@article{li2022asynchronous,
  title={Asynchronous Spatio-Temporal Memory Network for Continuous Event-Based Object Detection},
  author={Li, Jianing and Li, Jia and Zhu, Lin and Xiang, Xijie and Huang, Tiejun and Tian, Yonghong},
  journal={IEEE Transactions on Image Processing},
  volume={31},
  pages={2975--2987},
  year={2022},
  doi={10.1109/TIP.2022.3162962}
}

@article{chen2019multi,
  title={Multi-Cue Event Information Fusion for Pedestrian Detection with Neuromorphic Vision Sensors},
  author={Chen, Guang and Cao, Hu and Ye, Chao and Zhang, Zhen and Liu, Xiaolin and Mo, Xuhui and Qu, Zhen and Conradt, Jorg and R{\"o}hrbein, Florian and Knoll, Alois},
  journal={Frontiers in Neurorobotics},
  volume={13},
  pages={10},
  year={2019},
  doi={10.3389/fnbot.2019.00010}
}

@inproceedings{gehrig2023recurrent,
  title={Recurrent Vision Transformers for Object Detection with Event Cameras},
  author={Gehrig, Mathias and Scaramuzza, Davide},
  booktitle={Proceedings of the IEEE/CVF Conference on Computer Vision and Pattern Recognition},
  pages={13884--13893},
  year={2023}
}

@inproceedings{peng2023better,
  title={Better and Faster: Adaptive Event Conversion for Event-Based Object Detection},
  author={Peng, Yansong and Zhang, Yueyi and Xiong, Zhiwei and Sun, Xiaoyan and Wu, Feng},
  booktitle={Proceedings of the AAAI Conference on Artificial Intelligence},
  volume={37},
  number={2},
  pages={2056--2064},
  year={2023},
  doi={10.1609/aaai.v37i2.25300}
}

@inproceedings{peng2024scene,
  title={Scene adaptive sparse transformer for event-based object detection},
  author={Peng, Yansong and Li, Hebei and Zhang, Yueyi and Sun, Xiaoyan and Wu, Feng},
  booktitle={Proceedings of the IEEE/CVF Conference on Computer Vision and Pattern Recognition},
  pages={16794--16804},
  year={2024}
}

@inproceedings{zubic2024state,
  title={State space models for event cameras},
  author={Zubic, Nikola and Gehrig, Mathias and Scaramuzza, Davide},
  booktitle={Proceedings of the IEEE/CVF conference on computer vision and pattern recognition},
  pages={5819--5828},
  year={2024}
}

@inproceedings{wu2024leod,
  title={{LEOD}: Label-Efficient Object Detection for Event Cameras},
  author={Wu, Ziyi and Gehrig, Mathias and Lyu, Qing and Liu, Xudong and Gilitschenski, Igor},
  booktitle={Proceedings of the IEEE/CVF Conference on Computer Vision and Pattern Recognition},
  pages={16933--16943},
  year={2024}
}

@inproceedings{yang2025smamba,
  title={{SMamba}: Sparse Mamba for Event-Based Object Detection},
  author={Yang, Nan and Wang, Yang and Liu, Zhanwen and Li, Meng and An, Yisheng and Zhao, Xiangmo},
  booktitle={Proceedings of the AAAI Conference on Artificial Intelligence},
  volume={39},
  number={9},
  pages={9229--9237},
  year={2025},
  doi={10.1609/aaai.v39i9.32999}
}

@inproceedings{messikommer2020event,
  title={Event-Based Asynchronous Sparse Convolutional Networks},
  author={Messikommer, Nico and Gehrig, Daniel and Loquercio, Antonio and Scaramuzza, Davide},
  booktitle={European Conference on Computer Vision},
  year={2020}
}

@inproceedings{cordone2022object,
  title={Object Detection with Spiking Neural Networks on Automotive Event Data},
  author={Cordone, Loic and Miramond, Benoit and Thierion, Philippe},
  booktitle={International Joint Conference on Neural Networks},
  year={2022}
}

@misc{hao2026ssla,
  title={Low-Latency Event-Based Object Detection with Spatially-Sparse Linear Attention},
  author={Hao, Haiqing and Sui, Zhipeng and Zou, Rong and Dai, Zijia and Zubic, Nikola and Scaramuzza, Davide and Wang, Wenhui},
  year={2026},
  eprint={2603.06228},
  archivePrefix={arXiv},
  primaryClass={cs.CV}
}

@misc{zhang2024automotive,
  title={Automotive Object Detection via Learning Sparse Events by Spiking Neurons},
  author={Zhang, Hu and Li, Yanchen and Leng, Luziwei and Che, Kaiwei and Liu, Qian and Guo, Qinghai and Liao, Jianxing and Cheng, Ran},
  year={2024},
  eprint={2307.12900},
  archivePrefix={arXiv},
  primaryClass={cs.CV}
}

@inproceedings{mitrokhin2020learning,
  title={Learning Visual Motion Segmentation Using Event Surfaces},
  author={Mitrokhin, Anton and Hua, Zhiyuan and Fermuller, Cornelia and Aloimonos, Yiannis},
  booktitle={Proceedings of the IEEE/CVF Conference on Computer Vision and Pattern Recognition},
  pages={14414--14423},
  year={2020}
}

@inproceedings{alonso2019evsegnet,
  title={{EV-SegNet}: Semantic Segmentation for Event-Based Cameras},
  author={Alonso, Inigo and Murillo, Ana C.},
  booktitle={Proceedings of the IEEE/CVF Conference on Computer Vision and Pattern Recognition Workshops},
  year={2019}
}

@inproceedings{sun2022ess,
  title={{ESS}: Learning Event-Based Semantic Segmentation from Still Images},
  author={Sun, Zhaoning and Messikommer, Nico and Gehrig, Daniel and Scaramuzza, Davide},
  booktitle={European Conference on Computer Vision},
  pages={341--357},
  year={2022},
  organization={Springer}
}

@article{zhang2021exploring,
  title={Exploring Event-Driven Dynamic Context for Accident Scene Segmentation},
  author={Zhang, Jiqing and Yang, Xiaohan and Stiefelhagen, Rainer},
  journal={IEEE Transactions on Intelligent Transportation Systems},
  year={2021}
}

@inproceedings{das2024halsie,
  title={{HALSIE}: Hybrid Approach to Learning Segmentation by Simultaneously Exploiting Image and Event Modalities},
  author={Das Biswas, Soumyajit and Kosta, Adarsha and Liyanagedera, Chamika and Apolinario, Matthew and Roy, Kaushik},
  booktitle={Proceedings of the IEEE/CVF Winter Conference on Applications of Computer Vision},
  pages={5964--5974},
  year={2024}
}

@inproceedings{li2025efficient,
  title={Efficient Event-Based Semantic Segmentation via Exploiting Frame-Event Fusion: A Hybrid Neural Network Approach},
  author={Li, Kai and Zhao, Yucheng and Lyu, Gengyu and Deng, Yongjian},
  booktitle={Proceedings of the AAAI Conference on Artificial Intelligence},
  year={2025}
}

@article{zhang2023cmx,
  title={{CMX}: Cross-Modal Fusion for RGB-X Semantic Segmentation with Transformers},
  author={Zhang, Jiaming and Liu, Huayao and Yang, Kailun and Hu, Xinxin and Liu, Risheng and Stiefelhagen, Rainer},
  journal={IEEE Transactions on Intelligent Transportation Systems},
  volume={24},
  number={12},
  pages={14679--14694},
  year={2023},
  doi={10.1109/TITS.2023.3300537}
}

@inproceedings{zhang2023delivering,
  title={Delivering Arbitrary-Modal Semantic Segmentation},
  author={Zhang, Jiaming and Liu, Huayao and Yang, Kailun and Hu, Xinxin and Liu, Risheng and Stiefelhagen, Rainer},
  booktitle={Proceedings of the IEEE/CVF Conference on Computer Vision and Pattern Recognition},
  pages={1136--1147},
  year={2023}
}

@inproceedings{gu2026mambaseg,
  title={{MambaSeg}: Harnessing Mamba for Accurate and Efficient Image-Event Semantic Segmentation},
  author={Gu, Fuqiang and Li, Yuanke and Long, Xianlei and Ji, Kangping and Chen, Chao and Gu, Qingyi and Ni, Zhenliang},
  booktitle={Proceedings of the AAAI Conference on Artificial Intelligence},
  year={2026}
}

@article{sun2025dcr,
  title={{DCR-EFlow}: Dynamic Correlation Recurrent Architecture for Optical Flow Estimation Based on Event Cameras},
  author={Sun, Fanzhe and Su, Changqing and Xiong, Bo and Wang, Yu},
  journal={Intelligent Computing},
  volume={4},
  pages={0243},
  year={2025},
  doi={10.34133/icomputing.0243}
}

@article{wan2022learning,
  title={Learning dense and continuous optical flow from an event camera},
  author={Wan, Zhexiong and Dai, Yuchao and Mao, Yuxin},
  journal={IEEE Transactions on Image Processing},
  volume={31},
  pages={7237--7251},
  year={2022},
  publisher={IEEE}
}

@inproceedings{gehrig2021eraft,
  title={{E-RAFT}: Dense Optical Flow from Event Cameras},
  author={Gehrig, Mathias and Millh{\"a}usler, Mario and Gehrig, Daniel and Scaramuzza, Davide},
  booktitle={International Conference on 3D Vision},
  pages={197--206},
  year={2021},
  doi={10.1109/3DV53792.2021.00030}
}

@inproceedings{liu2023tma,
  title={{TMA}: Temporal Motion Aggregation for Event-based Optical Flow},
  author={Liu, Haotian and Chen, Guang and Qu, Sanqing and Zhang, Yanping and Li, Zhijun and Knoll, Alois and Jiang, Changjun},
  booktitle={Proceedings of the IEEE/CVF International Conference on Computer Vision},
  pages={9685--9694},
  year={2023}
}

@article{ye2025eva,
  title={Towards Anytime Optical Flow Estimation with Event Cameras},
  author={Ye, Yaozu and Shi, Hao and Yang, Kailun and Wang, Ze and Yin, Xiaoting and Lin, Yining and Liu, Mao and Wang, Yaonan and Wang, Kaiwei},
  journal={Sensors},
  volume={25},
  number={10},
  pages={3158},
  year={2025},
  doi={10.3390/s25103158}
}

@inproceedings{shiba2023taming,
  title={Secrets of Event-Based Optical Flow},
  author={Shiba, Shintaro and Aoki, Yoshimitsu and Gallego, Guillermo},
  booktitle={European Conference on Computer Vision},
  pages={628--645},
  year={2022}
}

@inproceedings{ding2022spatio,
  title={Spatio-Temporal Recurrent Networks for Event-Based Optical Flow Estimation},
  author={Ding, Ziluo and Zhao, Rui and Zhang, Jiyuan and Gao, Tianxiao and Xiong, Ruiqin and Yu, Zhaofei and Huang, Tiejun},
  booktitle={Proceedings of the AAAI Conference on Artificial Intelligence},
  volume={36},
  number={1},
  pages={525--533},
  year={2022}
}

@inproceedings{lee2020spikeflownet,
  title={{Spike-FlowNet}: Event-Based Optical Flow Estimation with Energy-Efficient Hybrid Neural Networks},
  author={Lee, Chankyu and Kosta, Adarsh Kumar and Zhu, Alex Zihao and Chaney, Kenneth and Daniilidis, Kostas and Roy, Kaushik},
  booktitle={European Conference on Computer Vision},
  pages={366--382},
  year={2020}
}

@article{tang2025event,
  title={Event Camera-Based Human Pose Estimation via Hybrid Spiking-Point Cloud Neural Architecture},
  author={Tang, Sichao and Lv, Hengyi and Li, Xiangzhi and Zhao, Yuchen and Zhang, Yisa and Feng, Yang},
  journal={Results in Engineering},
  volume={27},
  pages={106771},
  year={2025},
  doi={10.1016/j.rineng.2025.106771}
}

@inproceedings{tian2022etflownet,
  title={Event Transformer FlowNet for Optical Flow Estimation},
  author={Tian, Yi and Andrade-Cetto, Juan},
  booktitle={British Machine Vision Conference},
  year={2022}
}

@inproceedings{wu2022video,
  title={Video Interpolation by Event-Driven Anisotropic Adjustment of Optical Flow},
  author={Wu, Song and You, Kaichao and He, Weihua and Yang, Chen and Tian, Yang and Wang, Yaoyuan and Zhang, Ziyang and Liao, Jianxing},
  booktitle={European Conference on Computer Vision},
  pages={267--283},
  year={2022}
}

@inproceedings{scheerlinck2020fast,
  title={Fast Image Reconstruction with an Event Camera},
  author={Scheerlinck, Cedric and Rebecq, Henri and Gehrig, Daniel and Barnes, Nick and Mahony, Robert and Scaramuzza, Davide},
  booktitle={Proceedings of the IEEE/CVF Winter Conference on Applications of Computer Vision},
  pages={156--163},
  year={2020},
  doi={10.1109/WACV45572.2020.9093366}
}

@inproceedings{paredes2021back,
  title={Back to Event Basics: Self-Supervised Learning of Image Reconstruction for Event Cameras via Photometric Constancy},
  author={Paredes-Valles, Federico and de Croon, Guido C. H. E.},
  booktitle={Proceedings of the IEEE/CVF Conference on Computer Vision and Pattern Recognition},
  pages={3446--3455},
  year={2021},
  doi={10.1109/CVPR46437.2021.00345}
}

@inproceedings{weng2021event,
  title={Event-Based Video Reconstruction Using Transformer},
  author={Weng, Wenming and Zhang, Yueyi and Xiong, Zhiwei},
  booktitle={Proceedings of the IEEE/CVF International Conference on Computer Vision},
  pages={2563--2572},
  year={2021}
}

@inproceedings{han2022event,
  title={Event-Based Fusion for Motion Deblurring with Cross-Modal Attention},
  author={Sun, Lei and Sakaridis, Christos and Liang, Jingyun and Jiang, Qi and Yang, Kailun and Sun, Peng and Ye, Yaozu and Wang, Kaiwei and Van Gool, Luc},
  booktitle={European Conference on Computer Vision},
  pages={412--428},
  year={2022},
  organization={Springer},
  doi={10.1007/978-3-031-19797-0_24}
}

@inproceedings{yang2023event,
  title={Event-Based Motion Deblurring with Modality-Aware Decomposition and Recomposition},
  author={Yang, Wen and Wu, Jinjian and Li, Leida and Dong, Weisheng and Shi, Guangming},
  booktitle={Proceedings of the 31st ACM International Conference on Multimedia},
  pages={8327--8335},
  year={2023},
  doi={10.1145/3581783.3612505}
}

@inproceedings{sun2024motion,
  title={Motion Aware Event Representation-Driven Image Deblurring},
  author={Sun, Zhijing and Fu, Xueyang and Huang, Longzhuo and Liu, Aiping and Zha, Zheng-Jun},
  booktitle={European Conference on Computer Vision},
  pages={418--435},
  year={2024},
  organization={Springer},
  doi={10.1007/978-3-031-72952-2_24}
}

@inproceedings{xu2025motion,
  title={Motion-Adaptive Transformer for Event-Based Image Deblurring},
  author={Xu, Senyan and Sun, Zhijing and Zhong, Mingchen and Cao, Chengzhi and Liu, Yidi and Fu, Xueyang and Chen, Yan},
  booktitle={Proceedings of the AAAI Conference on Artificial Intelligence},
  volume={39},
  number={7},
  pages={6952--6960},
  year={2025},
  doi={10.1609/aaai.v39i7.32967}
}

@inproceedings{kim2024frequency,
  title={Frequency-Aware Event-Based Video Deblurring for Real-World Motion Blur},
  author={Kim, Taewoo and Cho, Hoonhee and Yoon, Kuk-Jin},
  booktitle={Proceedings of the IEEE/CVF Conference on Computer Vision and Pattern Recognition},
  pages={24966--24976},
  year={2024}
}

@inproceedings{lee2025distile2d,
  title={Distil-E2D: Distilling Image-to-Depth Priors for Event-Based Monocular Depth Estimation},
  author={Lee, Jie Long and Lee, Gim Hee},
  booktitle={Advances in Neural Information Processing Systems},
  year={2025}
}

@inproceedings{zhu2025depth,
  title={Depth Any Event Stream: Enhancing Event-based Monocular Depth Estimation via Dense-to-Sparse Distillation},
  author={Zhu, Jinjing and Pan, Tianbo and Cao, Zidong and Liu, Yexin and Kwok, James T. and Xiong, Hui},
  booktitle={Proceedings of the IEEE/CVF International Conference on Computer Vision},
  pages={5146--5155},
  year={2025}
}

@inproceedings{hidalgo2020learning,
  title={Learning Monocular Dense Depth from Events},
  author={Hidalgo-Carri{\'o}, Javier and Gehrig, Daniel and Scaramuzza, Davide},
  booktitle={2020 International Conference on 3D Vision (3DV)},
  pages={534--542},
  year={2020},
  organization={IEEE}
}

@article{gehrig2021ramnet,
  title={Combining Events and Frames Using Recurrent Asynchronous Multimodal Networks for Monocular Depth Prediction},
  author={Gehrig, Daniel and R{\"u}egg, Michelle and Gehrig, Mathias and Hidalgo-Carri{\'o}, Javier and Scaramuzza, Davide},
  journal={IEEE Robotics and Automation Letters},
  volume={6},
  number={2},
  pages={2822--2829},
  year={2021},
  publisher={IEEE}
}

@article{sabater2023event,
  title={Event Transformer+. A Multi-Purpose Solution for Efficient Event Data Processing},
  author={Sabater, Alberto and Montesano, Luis and Murillo, Ana C.},
  journal={IEEE Transactions on Pattern Analysis and Machine Intelligence},
  volume={45},
  number={12},
  pages={16013--16020},
  year={2023},
  publisher={IEEE},
  doi={10.1109/TPAMI.2023.3311336}
}

@inproceedings{devulapally2024multi,
  title={Multi-Modal Fusion of Event and RGB for Monocular Depth Estimation Using a Unified Transformer-Based Architecture},
  author={Devulapally, Anusha and Khan, Md Fahim Faysal and Advani, Siddharth and Narayanan, Vijaykrishnan},
  booktitle={Proceedings of the IEEE/CVF Conference on Computer Vision and Pattern Recognition Workshops},
  pages={2081--2089},
  year={2024}
}

@inproceedings{pan2024srfnet,
  title={SRFNet: Monocular Depth Estimation with Fine-Grained Structure via Spatial Reliability-Oriented Fusion of Frames and Events},
  author={Pan, Tianbo and Cao, Zidong and Wang, Lin},
  booktitle={2024 IEEE International Conference on Robotics and Automation (ICRA)},
  pages={10695--10702},
  year={2024},
  organization={IEEE}
}

@article{liu2022eventdepth,
  title={Event-Based Monocular Dense Depth Estimation with Recurrent Transformers},
  author={Liu, Xueyi and Li, Jia and Fan, Xingxing and Tian, Yonghong},
  journal={arXiv preprint arXiv:2212.02791},
  year={2022}
}

@inproceedings{zhu2023selfsupervised,
  title={Self-Supervised Event-Based Monocular Depth Estimation Using Cross-Modal Consistency},
  author={Zhu, Junyu and Liu, Lina and Jiang, Bofeng and Wen, Feng and Zhang, Hongbo and Li, Wanlong and Liu, Yong},
  booktitle={2023 IEEE/RSJ International Conference on Intelligent Robots and Systems (IROS)},
  pages={7704--7710},
  year={2023},
  organization={IEEE},
  doi={10.1109/IROS55552.2023.10342434}
}

@article{gehrig2021dsec,
  title={DSEC: A Stereo Event Camera Dataset for Driving Scenarios},
  author={Gehrig, Mathias and Aarents, Willem and Gehrig, Daniel and Scaramuzza, Davide},
  journal={IEEE Robotics and Automation Letters},
  volume={6},
  number={3},
  pages={4947--4954},
  year={2021},
  publisher={IEEE}
}

@inproceedings{li2024coarse,
  title={A Coarse-to-Fine Fusion Network for Event-Based Image Deblurring},
  author={Li, Huan and Shi, Hailong and Gao, Xingyu},
  booktitle={Proceedings of the Thirty-Third International Joint Conference on Artificial Intelligence},
  pages={953--961},
  year={2024},
  doi={10.24963/ijcai.2024/108}
}

@inproceedings{zhou2023rgb,
  title={{RGB}-Event Fusion for Moving Object Detection in Autonomous Driving},
  author={Zhou, Zhuyun and Wu, Zongwei and Boutteau, R{\'e}mi and Yang, Fan and Demonceaux, C{\'e}dric and Ginhac, Dominique},
  booktitle={Proceedings of the IEEE International Conference on Robotics and Automation},
  year={2023},
  organization={IEEE}
}

@inproceedings{li2019event,
  title={Event-Based Vision Enhanced: A Joint Detection Framework in Autonomous Driving},
  author={Li, Jianing and Dong, Siwei and Yu, Zhaofei and Tian, Yonghong and Huang, Tiejun},
  booktitle={Proceedings of the IEEE International Conference on Multimedia and Expo},
  year={2019},
  organization={IEEE}
}

\end{document}